\documentclass[11pt]{aastex}
\usepackage{graphicx,emulateapj5,apjfonts}
\usepackage{onecolfloat}
\usepackage{epsf}

\shortauthors{Bell et al.}
\shorttitle{The Evolution of Early-Type Galaxies}

\newcommand{\combo}{{\rm COMBO-17 }}
\newcommand{\peg}{{\sc P\'egase }}

\slugcomment{{\sc To appear in the Astrophysical Journal: } 20 June 2004}

\begin{document}


\def\head{

\title{Nearly 5000 Distant Early-Type Galaxies in COMBO-17:\\
a Red Sequence and its Evolution since $z \sim 1$}

\author{Eric F.\ Bell$^1$, Christian Wolf$^2$, Klaus Meisenheimer$^1$, 
Hans-Walter Rix$^1$, Andrea Borch$^1$, Simon Dye$^3$, 
Martina Kleinheinrich$^1$, Lutz Wisotzki$^4$, and Daniel H.\ McIntosh$^5$}
\affil{$^1$ Max Planck Institut f\"ur Astronomie,
K\"onigstuhl 17, D-69117 Heidelberg, Germany; 
\texttt{bell,meise,rix,borch,martina@mpia.de} \\ 
$^2$ Department of Physics, Denys Wilkinson Bldg., University of 
Oxford, Keble Road, Oxford, OX1 3RH, UK; 
\texttt{cwolf@astro.ox.ac.uk} \\
$^3$ Astrophysics Group, Blackett Laboratory, Imperial College, 
Prince Consort Road, London SW7 2BW, UK; \texttt{s.dye01@imperial.ac.uk} \\
$^4$ Astrophysikalisches Institut Potsdam, An der Sternwarte 16, 
D-14482 Potsdam, Germany;  \texttt{lwisotzki@aip.de} \\
$^5$ Department of Astronomy, University of Massachusetts, 
710 North Pleasant Street,
Amherst, MA 01003-9305; \texttt{dmac@hamerkop.astro.umass.edu}
}

\begin{abstract}
We present the rest-frame colors and luminosities of $\sim 25000$ 
$m_R \la 24$ galaxies in the redshift range $0.2 < z \le 1.1$, 
drawn from 0.78 square degrees of the \combo survey.  We find
that the rest-frame color distribution of these galaxies is
bimodal at all redshifts out to $z \sim 1$.  This bimodality
permits a model-independent definition of red, early-type 
galaxies and blue, late-type galaxies at any given redshift.
The colors of the blue peak become redder towards the present day, and 
the number density of blue luminous galaxies has dropped
strongly since $z \sim 1$.
Focusing on the red galaxies, we find that they 
populate a color-magnitude relation.  Such red 
sequences have been identified in galaxy cluster environments, but 
our data show that such a sequence exists over this redshift range
even when averaging over all environments.
The mean color of the red galaxy sequence
evolves with redshift in a way that is consistent with
the aging of an ancient stellar population.  
The rest-frame $B$-band luminosity density in red galaxies
evolves only mildly with redshift in a $\Lambda$-dominated
cold dark matter universe.  Accounting for the change in 
stellar mass-to-light ratio implied by 
the redshift evolution in red galaxy colors,
the \combo data indicate an increase in stellar mass on the red sequence
by a factor of two since $z \sim 1$.  The largest source
of uncertainty is large-scale structure, 
implying that considerably
larger surveys are necessary to further refine this result.
We explore mechanisms that may drive this evolution 
in the red galaxy population, finding that both galaxy merging and truncation 
of star formation in some fraction of the blue, star-forming population 
are required to fully explain the properties
of these galaxies.
\end{abstract}

\keywords{galaxies: evolution --- galaxies: general --- 
galaxies: luminosity function --- galaxies: elliptical and lenticular ---
galaxies: stellar content --- surveys }
}

\twocolumn[\head]

\section{Introduction}

The evolution of early-type galaxies since early in the Universe's history 
is a highly active area of research.  The current world model
for the evolution of the Universe is the Cold Dark Matter
paradigm; the incarnation which appears consistent 
with most observations adopts $\Omega_{\rm m} = 0.3$, 
$\Omega_{\Lambda} = 0.7$, and $H_0 = 100\,h$\,km\,s$^{-1}$\,Mpc$^{-1}$,
where $h \sim 0.7$ \citep[the $\Lambda$CDM 
paradigm; e.g.,][]{freedman01,efstathiou02,pryke02,spergel03}.  
In this model, early-type galaxies form hierarchically 
through mergers of pre-existing galaxies 
\citep[e.g.,][]{white91,barnes92,cole00}.
A generic prediction of this type of model is an
increase in the stellar mass density of 
the early-type galaxy population since $z \sim 1$, 
as large galaxies are assembled primarily at relatively
late times \citep[e.g.,][]{aragon98,kauffmann98}.
Therefore, an important test of the hierarchical formation
scenario is to quantify the evolution of early-type galaxies
since $z \sim 1$.  
In this paper, we discuss the color distributions of 
a sample of $\sim 25000$ galaxies from the COMBO-17 survey
\citep[`Classifying Objects by Medium-Band 
Observations in 17 Filters';][]{wolf03}, 
focusing on an objective definition of early-type galaxies at
all redshifts.  We then use this objectively-defined sample of
early-type galaxies with $0.2 < z \le 1.1$ to examine the evidence
for a substantial build-up of stellar mass in red galaxies since $z \sim 1$.

In the local Universe, early-type galaxies have historically 
been defined in terms of morphology.  Loosely speaking, 
galaxies with dominant smooth spheroidal components, and 
at most a modest and smooth disk, will be 
classified as an `early-type' galaxy.  These galaxies tend to be 
red in optical color both locally \citep[e.g.,][]{schweizer92}
and out to $z \sim 1$ \citep[e.g.,][]{kodama99}.
This definition of
early-type galaxies is being challenged by the recent 
demonstration that galaxies are distributed in color
space in a bimodal distribution
\citep[see, e.g.,][]{strateva01,hogg02,colblanton}.
One peak is red, and consists
mostly of non-star-forming galaxies earlier than {\it Sa} in morphological 
type.  The other peak is blue
and consists primarily of star-forming galaxies later than {\it Sb} in 
morphological type \citep{strateva01}.  Because of this bimodality
in color and therefore star formation history (SFH), it is
defensible, and perhaps more natural, to define early types in terms of
colors (i.e., spectral energy distribution or SFH).  
These different definitions seem largely 
consistent \citep[e.g.,][]{schweizer92,strateva01,bell03,bell04}.
In this paper, we split the galaxy population into 
red and blue subsets based on their rest-frame optical colors, 
exploring the evolution of these populations with cosmic time.

In the local Universe, the blue-peak galaxies (later types)
show a scattered but systematic variation 
of color with magnitude, in the sense
that luminous galaxies tend to be somewhat less blue
\citep[e.g.,][]{tully98,hogg02}.  These color changes, 
when examined in detail, are due to changes in the mean ages, metallicities,
and dust contents
of galaxies with galaxy magnitude, such that brighter galaxies
tend to be older, dustier, and more metal-rich 
\citep[see, e.g.,][]{tully98,bdj}.
In contrast, red-peak galaxies (earlier types) 
have a tighter and well-defined relationship between color
and magnitude, such that bright galaxies are typically 
redder \citep[e.g.,][]{sandage78,ble92,schweizer92,terlevich01}.  
This color-magnitude relation (CMR) is well-established
in overdense and cluster environments, where early-type galaxies
are much more common \citep{dressler80,dressler97}.  Nevertheless, the CMR 
also exists among present-day field early-type 
galaxies \citep[e.g.,][]{sandage78,schweizer92,hogg03}.
The scatter, slope and evolution of the CMR 
in clusters of galaxies are all consistent with 
the interpretation of the CMR as primarily
a metallicity sequence of old galaxies, where more massive
galaxies are more metal-rich 
\citep[e.g.,][]{ble92,kodama97,vazdekis01,bernardi03}.
A modest age spread is possible \citep{trager00}. 

The evolution of the properties of individual early-type galaxies
with redshift has been studied in clusters of galaxies, where the assembly of 
sizeable galaxy samples has been feasible.  It appears 
as if individual early-type galaxies were formed relatively
quickly at high redshift and simply aged to the present
day through so-called passive evolution 
\citep[e.g.,][]{kodama97,kelson01,vandokkum03}.  This has been used
to justify `monolithic collapse' models of early-type galaxy
formation, where these galaxies form quickly early in the history
of the Universe, and simply age to the present day 
\citep[e.g.,][]{els,larson75,kodama97,nulsen97}.  Yet,
monolithic collapse, at best, is mildly unphysical;
the hierarchical build-up of galaxies through mergers
and interactions must happen at some level
\citep[e.g.][]{white91}.  

An elegant interpretation of the apparent contradiction
between seemingly passive evolution and the hierarchical 
build-up of early-type galaxies has been put forward
by \citet[progenitor bias]{vandokkum01}.  They note that
studies of the properties of early-type
galaxies in isolation (rightfully) fail to study the star-forming
precursors of galaxies that are later destined to become 
early types.  Thus, individual early-type galaxies appear old at all 
redshifts regardless of whether they were formed in a burst
at redshift infinity or whether their formation happens 
through violent mergers which consume all of the 
gas in the interacting galaxies, leaving a spheroidal 
remnant \citep[e.g.,][]{toomre,barnes96}.  This is a common 
theme in many contemporary studies of early-type 
galaxies; the distinction between when the {\it stars} in 
early-type galaxies were formed, and when the
{\it galaxies} were assembled into recognizable entities.
Studies of galaxies
selected by strong gravitational lensing largely avoid 
this bias by selecting by mass 
\citep[e.g.,][]{kochanek00,rusin03,vandeven03}; yet,
these studies also find rather ancient-looking stellar populations.

Thus, a more important constraint on the evolution 
of early-type galaxies --- and therefore, a stronger test of 
galaxy formation models --- is the evolution of their number density
and luminosity function (whether the early-type galaxy population is
defined by morphology or color).
Here, the situation is somewhat less
clear.  Many studies find that the evolution of the early-type
population is close to passive up to 
$z \sim 1$, with perhaps a slight decrease in number 
density above $z \sim 1$ \citep[e.g.,][]{lilly95,schade99,lin99,
cimatti02,firth02,im02,chen03,pozetti03}.
Other studies find relatively rapid evolution. 
For example, \citet{kauffmann96}
find number evolution of a factor of three by $z \sim 1$, 
although \citet{totani98} argue that spectroscopic incompleteness causes
this apparent evolution, and state that Kauffmann et al's
results are consistent with passive evolution.
\citet{wolf03} find a rapid increase
in the luminosity density of galaxies with the colors
of {\it present-day} early-type galaxies from $z \sim 1$ until 
the present day; yet, passive evolution
and reddening of the galaxy population is likely responsible
for much of this.

Nevertheless, a problem shared by most surveys of early-type galaxies
at $z \ga 0.5$ has been susceptibility to large
scale structure.  Surveys to date have typically had samples of
only $50-1500$ early-type galaxies
\citep[e.g.,][]{kauffmann98,schade99,lin99,firth02,chen03,pozetti03}.
Furthermore, these samples are usually from relatively small 
areas on the sky \citep[the largest are CNOC2 and the LCIRS with 1400 
square arcminutes each;][]{lin99,chen03}.  Worse still, early-type 
galaxies are strongly clustered 
\citep[e.g.,][]{dressler80,dressler97,moustakas02,daddi02}, 
which means that the measured 
number evolution is particularly
sensitive to large-scale structure.  Thus, to accurately measure the 
evolution of the stellar mass in early-type galaxies, 
it is necessary to use large samples drawn from large areas 
on the sky.\footnote{The cosmic variance error decreases 
by a factor of two for more than an order of magnitude increase in 
comoving volume probed (if the area is contiguous), or 
for a factor of four increase in the number of independent pointings
\citep{somer04}. Therefore, to derive an accurate cosmic average
a survey with many independent pointings is ideal.  Typically, 
surveys cover a few independent pointings for practicality;
ours is no exception.}

In this paper, we bring the largest present-day sample
of galaxies with $0.2 < z \le 1.1$ to bear on this 
important problem.  The COMBO-17 survey has to date imaged and fully
analyzed three $\sim$0.25 square degree fields in 5 broad-band and 
12 narrow-band filters (data for a fourth field has been taken but
is not yet analyzed).  The 5 broad and 12 narrow-bands allow
the construction of photometric redshifts and rest-frame colors with 
extraordinary precision (one can think of the data as a $R \sim 10$ 
ultra-low resolution spectrum rather as a photometric dataset), yielding
galaxy redshifts accurate to a few percent for a total sample
of $\sim 25000$ galaxies in 0.78 square degrees (2800 square arcminutes)
with $m_R < 24$.  We use this sample to 
discuss the colors of galaxies with $0.2 < z \le 1.1$,
focusing particularly on the reddest galaxies at all redshifts.
We put forward an objective definition of early-type galaxies, and
use this definition to explore the  
evolution of the red galaxy population from redshift unity to the present day.

This paper is set out as follows.  The COMBO-17 survey and
the main sources of error are briefly discussed in \S \ref{sec:data}.
In \S \ref{sec:col}, we explore the colors of the entire COMBO-17 
galaxy population with $0.2 < z \le 1.1$.  Then, in \S \ref{sec:cmr}, 
we focus on the reddest galaxies at all redshifts, exploring their
CMR in more detail.  Defining early-type galaxies as galaxies on the CMR, 
we then go on in \S \ref{sec:lf} 
to explore the evolution of the luminosity function
of red galaxies.  We discuss the results in
\S \ref{sec:disc}, and present the conclusions in \S \ref{sec:conc}.
Throughout, we assume $\Omega_{\rm m} = 0.3$, 
$\Omega_{\Lambda} = 0.7$, and $H_0 = 100\,h$\,km\,s$^{-1}$\,Mpc$^{-1}$.
In order to estimate the passive evolution of stellar populations
as a function of lookback time, we adopt $h = 0.7$
to be consistent with the {\it Hubble Space Telescope}
Key Project distance scale \citep{freedman01} and 
the recent results from the {\it Wilkinson Microwave Anisotropy Probe} 
\citep{spergel03}.

\section{The Data} \label{sec:data}

The \combo data, the sample selection, redshift estimation,
construction of rest-frame luminosities, and completeness are
all discussed in detail by \citet[W03 hereafter]{wolf03}, 
\citet[W01 hereafter]{wolf01a}, \citet{wolf01b}, and in a forthcoming
technical paper (Wolf et al., in preparation).  Here,
we briefly discuss the relevant limitations and sources of 
uncertainty in the \combo data.

\subsection{Data and Sample Selection} 

To date, the \combo project has surveyed 
0.78 square degrees to deep limits 
in three pointings  
of the Wide Field Imager \citep[WFI;][]{baade98,baade99}
at the Max Planck Gesellschaft/Euro\-pe\-an Southern Observatory
2.2-m telescope at La Silla, Chile.  \combo targeted
the Chandra Deep Field South, an equatorial field, and 
a field centered on the $z=0.16$ Abell 901 cluster
(for the coordinates 
of the three fields, see W03).  We study the galaxy population
at $z \ge 0.2$, therefore the presence of the low-$z$ Abell 901
cluster of galaxies in one of the survey fields will not 
significantly bias our analysis.
The WFI has eight
2k$\times$4k CCDs, a field of view of $34\arcmin \times 33\arcmin$, 
and a pixel scale of $0\farcs238$/pixel.  A total of 
$\sim 160$ksec per field were taken in 5 wide and 12 medium
passbands with 3640{\AA}$ \le \lambda_{\rm obs} \le 9140${\AA};
an ultra-deep 20ksec $R$-band with seeing below
$0\farcs8$ and a $10\sigma$ limiting magnitude of 25.2, 
and deep $UBVI$ and 12 medium passbands designed to allow the 
construction of low-resolution $R\sim 10$ spectra, or 
equivalently accurate photometric redshift estimates, for
a total of $\sim 25000$ objects with $0.2 \le z \le 1.2$.
The comoving survey volume is $\sim 10^6 h^{-3}$\,Mpc$^3$ 
in the interval $0.2 < z \le 1.1$, split between three disjoint roughly 
equally-sized fields.

Galaxies were detected in the deep $R$-band frames using 
SExtractor \citep{sex}, and we adopt the {\sc mag-best} 
magnitude as our total magnitude estimate \citep[a][magnitude in uncrowded
regions]{kron}.  \citet{sex} show that 
this magnitude underestimates the flux in galaxies
by $\sim$6\%. We do not correct for this effect as almost
all local galaxy surveys adopt similarly biased 
Kron or Petrosian magnitudes \citep[e.g.,][]{blanton03a,skrut}. 
The spectral shapes 
for $R$-band detected objects were measured by 
performing seeing-adaptive, weighted aperture photometry
in all 17 frames at the position of the $R$-band
detected object using the package {\sc mpiaphot}
\citep[Meisenheimer et al., in preparation]{roser91}.
This package measures the peak surface brightness
of all of the images smoothed to identical seeing ($\sim 1\farcs5$) to 
maximize signal-to-noise, allowing one to construct the best possible
photometric redshifts. Thus, the colors are 
essentially total for distant objects and are central colors
for more extended, nearby galaxies.  The effects of this aperture
bias are mild, and are discussed later in \S \ref{sec:cmr}.
Photometric calibration was achieved using spectrophotometric
standards in each \combo field.  All magnitudes are quoted in 
Vega-normalized magnitudes.  The median galaxy in our sample with $m_R \sim 22$
has observed galaxy magnitude errors of 
$\delta_R \sim 0.01$ mag, $\delta_{B,V,I} \sim 0.05 $ mag, 
$\delta_U \sim 0.3$ mag, and 
$\delta_{\rm Medium}$ for the medium passbands 
increasing from 0.03 mag at the red end to 
0.2 mag at the blue end. 

\subsection{Photometric Redshifts, Spectral Templates and Rest-Frame Colors}

The full survey dataset is photometrically classified
into stars, galaxies and AGN using the 17-passband `fuzzy' spectrum (W01).
No attempt is made to use morphological information, as, for example, 
double stars may contaminate the galaxy catalog, and 
compact galaxies may contaminate the stellar or AGN catalogs.  
For galaxy classification, we use \peg \citep[see][for 
an earlier version of the model]{fioc97} model spectra.  
The template spectra are a two-dimensional age/reddening sequence, 
where a fixed exponential star formation timescale $\tau = 1$\,Gyr is
assumed, ages vary between 50\,Myr and 10\,Gyr and the 
reddening $E(B-V)$ can be as large as 0.5 mag, adopting a
Small Magellanic Cloud Bar extinction curve\footnote{This classifier uses
a different template set from W03, who use empirical templates from 
\citet{kinney96}.  This completely model-based (somewhat unphysical) 
template set results 
in superior photometric redshifts to the template set used by W03.  
Owing to the almost complete age/metallicity/dust degeneracy, 
we attach no significance to the physical parameters used to 
derive the new template set.}.
These spectra are defined in the 
interval 1216{\AA}$\le \lambda_{\rm obs} \le 3{\micron} $
for elliptical through to starburst galaxies.  Trustworthy templates
shortwards of 1216{\AA} have not yet been included, limiting redshift 
measurements
at this stage to $z < 1.40$.  For our subsequent analysis the 
redshift range is further limited to 
$z \le 1.1$ in order to have at least two filters 
redwards of the 4000{\AA} break.  The template library
does not evolve, and cannot perfectly describe spectral 
types which are highly unusual in the local Universe, such as 
post-starburst E$+$A galaxies.  Furthermore, low-luminosity
or type 2 AGN will be misclassified as galaxies, sometimes
leading to increased redshift errors.
For galaxies and AGN, galaxy redshifts
are estimated simultaneously with the template type (W01).  A description
of an earlier version of this methodology was presented by W03, 
and a full description and analysis of these photometric 
redshifts and classifications will be presented by 
Wolf et al.\ (in preparation).

The galaxy redshift estimate quality has been tested by comparison
with spectroscopic redshifts for many hundreds of galaxies (see
Wolf et al., in preparation).  The redshift quality depends primarily
on apparent magnitude, and is in excellent agreement with the simulations
of \citet{wolf01a}.  At bright limits $m_R < 20$, 
the redshifts are accurate to $\delta z / (1+z) \sim 0.01$, and the
error is dominated by mismatches between template and real galaxy
spectra.  At the median apparent magnitude $m_R \sim 22$, 
$\delta z / (1+z) \sim 0.02$.  For the faintest galaxies, redshift
accuracy approaches those achievable using traditional broad-band photometric
surveys, $\delta z/(1+z) \ga 0.05$.  
Restframe colors and luminosities are constructed on a 
galaxy-by-galaxy basis by $k$-correcting the nearest observed broad 
band flux.  To estimate this $k$-correction, we convolve the best-fit 
spectrum with 
the filter curves for the Johnson
$U$, $B$, and $V$ passbands.  
Typical rest-frame color accuracy is $\sim 0.1$ mag, 
corresponding to the redshift uncertainty $\delta z/(1+z) \sim 0.02$.
Absolute magnitudes
have an additional error of $\sim 0.1$ mag ($z \ga 0.5$) and
$\sim 0.2$ mag ($z \sim 0.3$) owing to distance uncertainties.

\begin{figure*}[tbh]
\hspace{0.0cm}
\epsfxsize=17cm
\epsfbox{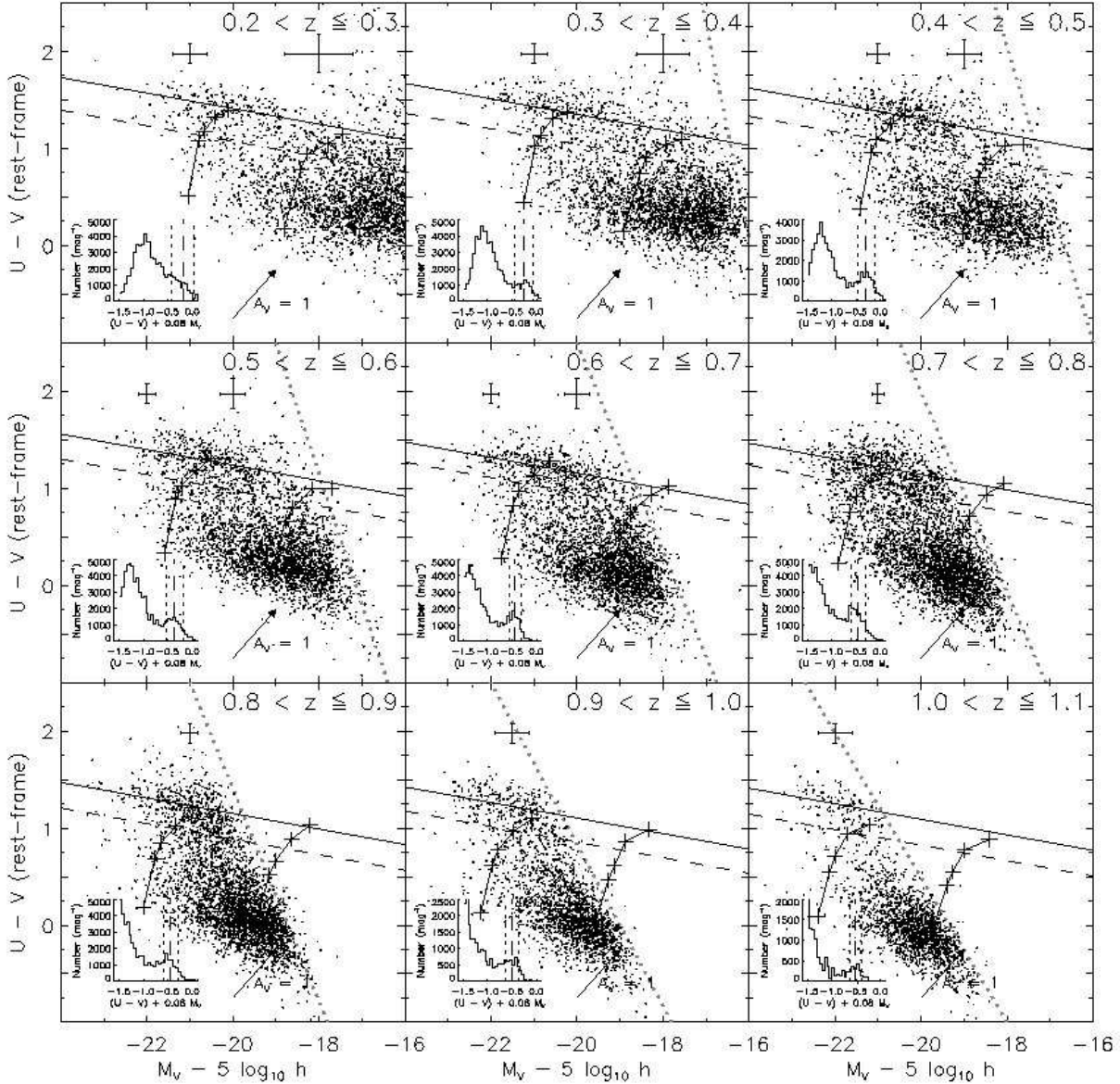}
\vspace{-0.4cm}
\caption{\label{fig:cmr} The rest-frame $U - V$ color of $\sim$25000 galaxies
against the absolute magnitude in $V$-band, $M_V - 5 \log_{10} h$.
We show the distribution of galaxies in nine different redshift bins.
A fit to the color-magnitude
relation of red-sequence galaxies with a 
fixed slope of $-0.08$ is shown by the solid
line, and the `Butcher-Oemler' style-cut between red and blue galaxies
is shown by the dashed line parallel to the early-type galaxy CMR
(see \S \ref{sec:lf}).  The sloping cutoff in the distribution of galaxies
at the faint end is due to the $m_R \la 24$ magnitude limit of the survey.
This cutoff cannot be exactly depicted owing to the color
dependence in completeness and the varying transformation
between observed $R$-band magnitude and rest-frame $V$-band magnitude; 
nevertheless, the dotted grey line shows very schematically the rough
completeness limit. 
The lines with crosses show the colors and magnitudes
of model galaxies with truncated SFHs at constant
stellar mass, described in more detail in \S \ref{sec:origin}.  Representative
error bars are shown.
For reference, we show a reddening vector from \citet{calzetti00} 
assuming $A_V = 1$ mag; a Milky-Way extinction curve gives the same 
vector.  Shown in the inset panels is the color distribution of the red
peak at each redshift 
(with the slope of the CMR taken out); the dashed line indicates
the position of the biweight mean that we adopt as the CMR ridge-line,
and the dotted lines show the biweight sigma that we adopt as the CMR
red sequence's scatter.   
 }
\vspace{-0.2cm}
\end{figure*}

\clearpage

\subsection{Completeness}

We include only galaxies with successful photometric redshift
and spectral classifications.  Therefore, when constructing
luminosity functions and densities, we must account for 
the sample completeness.  This is done using 
extensive Monte-Carlo simulations of the galaxy detection, 
photometry and classification process (W01); example completeness maps
are shown in Fig.\ 7 of W03.  For the red galaxies
which are the primary aim of this paper, the 90\% and 
50\% completeness limits are at roughly $m_R \sim 23.2$ and 
$\sim 23.8$, respectively.

\section{The Color Distribution of Galaxies with $0.2 < z \le 1.1$} 
  \label{sec:col}

One of the main aims of this paper is to explore and discuss 
the color distribution of galaxies in the interval $0.2 < z \le 1.1$.
We choose to quantify this in terms of $U-V$ color
as a function of $V$-band absolute magnitude, because
the $U-V$ color straddles the 4000{\AA} break and is therefore
particularly sensitive to age and metallicity variations
of the stellar populations in galaxies, and for 
consistency with well-known works from the literature
\citep[e.g.,][]{sandage78,ble92,schweizer92}.
In Fig.\ \ref{fig:cmr}, we show the distribution of rest-frame 
$U - V$ galaxy color against the $V$-band absolute magnitude
in nine different redshift bins covering $0.2 < z \le 1.1$.

Fig.\ \ref{fig:cmr} shows one of the key observational results of this paper,
that {\it the distribution
of galaxies in the color--magnitude diagram is 
bimodal at all redshifts out to $z \sim 1$}.  
\footnote{It is worth noting that the full range of rest-frame $U-V$ colors
is covered smoothly by the template spectra, making it 
unlikely that the algorithm artificially focuses galaxies
away from the gap between the red and blue galaxy sequences.
Furthermore, red-sequence galaxies at 
$z \sim 1$ have $U-V \sim 1.2$, and are the same color as 
gap galaxies at $z \sim 0.2$, arguing further against
artificial de-focusing of $U-V \sim 1.2$ galaxies
to redder or bluer colors.  }  
This confirms and extends 
9\,Gyr back in time the {\it Sloan Digital Sky Survey} (SDSS) results
from e.g., \citet{strateva01}, \citet{hogg02}, or \citet{colblanton},
who established and characterized the bimodality of the 
color distribution of galaxies in the local
Universe.  Furthermore, this confirms the suggestion of 
\citet{im02}, who saw hints of a bimodal color distribution 
out to $z \sim 1$.\footnote{The {\sc Deep2} survey with Keck 
also appears to find this bimodality (B.\ Weiner, 2003, priv. comm.)}  
For immediate low-redshift comparison, we show a 
synthesized $U-V$ color-magnitude
diagram for galaxies from the SDSS Early Data Release in 
the Appendix.  
We defer a fuller analysis of this bimodality to a future work, 
but do return to this issue briefly in \S \ref{sec:syn}.

Beyond the bimodality, it is important to notice two features of 
Fig.\ \ref{fig:cmr}.  Firstly, the rest-frame $U-V$ colors of blue 
galaxies become redder (at a given absolute magnitude)
$\Delta (U-V) \sim 0.5$ from $z \sim 1$ to the present day, 
probably indicating older mean ages and/or larger dust contents, perhaps
with a small contribution from metallicity evolution at a given luminosity.  
In addition, the number density of blue galaxies, especially
at high luminosities, significantly evolves.  
Because of the different volumes sampled by \combo in each 
redshift interval, this strong evolution in the number density
of luminous blue galaxies is not apparent in Fig.\ \ref{fig:cmr}.
W03 studied this issue in more detail; when volume-corrected,
the number density of 
faint star-forming galaxies (spectral Type 4) remains almost unchanged,
the abundance of luminous galaxies with starburst spectra 
drops precipitously from redshift unity to the present epoch
\citep[Fig.\ 16 of W03; also, e.g.,][]{cowie96}.  
This result, which is 
discussed much more by W03, also agrees
with the CNOC2 \citep{lin99} and CFRS \citep{lilly95} surveys,
who found strong evolution in the abundance of star-forming, blue galaxies
with redshift.  We choose not to elaborate further on the evolving
colors or luminosities of star-forming galaxies at this stage.

Secondly, bearing in mind the $\ga 0.1$ mag errors in $k$-corrected
rest-frame $U-V$ colors, the red galaxies form a relatively well-defined
sequence in the color-magnitude plane for $0.2 < z \le 1.1$.  
Furthermore, the mean color of this red sequence
evolves with redshift, in the sense
that galaxies become bluer at a fixed luminosity.
This evolution will be quantified in \S \ref{sec:cmr}.
Both observations agree with studies of 
galaxies in clusters at redshifts up to unity 
\citep[e.g.,][]{ble92,kodama97,terlevich01,vandokkum01}, 
but we extend these results by 
showing that they apply to volumes which average over many different
environments.
The red sequence has a significant slope in the color-magnitude
plane, therefore can be rightfully termed a color-magnitude relation
(CMR) in all redshift bins up to 
at least $z \sim 0.8$.  At $z \ga 0.8$, the magnitude range is so 
narrow that a constant color is also quite plausible.
It is the CMR of the red sequence galaxies that we study in detail 
in the rest of this paper.

\begin{figure}[tb]
\vspace{-0.5cm}
\hspace{-0.5cm}
\epsfbox{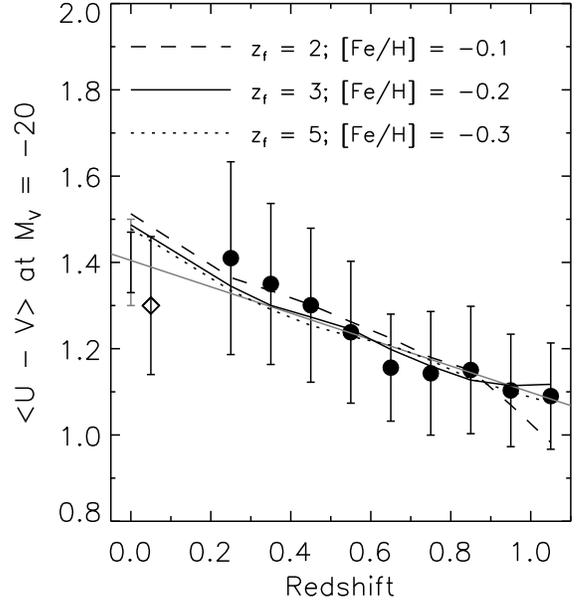}
\vspace{-0.2cm}
\caption{\label{fig:cmrint} Color evolution of the `red sequence', 
represented by the intercept of the CMR fits at
$M_V - 5 \log_{10} h = -20$ (solid circles) as a function
of redshift.  The error bars show the 
biweight scatter in each redshift bin.  These highly conservative errors
are much larger than the formal errors, and likely overestimate
the systematic errors from calibration, template/galaxy SED mismatches
and small redshift focusing issues, which we expect to be less
than 0.1 mag in $U-V$ color.  The lines show
the expected color evolution of single-age stellar populations with different
formation redshifts and metallicities, as given in the Figure legend.
The diamond with error bars (offset from zero redshift for clarity)
shows the $z = 0$ $U-V$ CMR zero point 
synthesized from SDSS $ugr$ data analyzed in the same way as these
data (see the Appendix for details).  The naked error bars show
the $U-V$ CMR zero point for the 
Abell 754 galaxy cluster \citep{mcintosh03}, which is also consistent 
with the Nearby Field Galaxy Survey \citep{jansen00} 
early-type galaxy CMR.  The black 
naked error bar shows the Abell 754 scatter around the CMR, and 
the expected systematic uncertainty is shown by a grey error bar. 
The grey line shows a fit to all of the data (including the Abell 754
and SDSS colors).
  } \vspace{-0.2cm}
\end{figure}

\clearpage

\section{The Color-Magnitude Relation of Red Galaxies} \label{sec:cmr}

We now quantify the redshift evolution of the CMR 
by examining its zero point.  In order to 
ensure consistency between the zero point estimates 
for different bins, we constrain the slope of 
the CMR to the locally-determined value for cluster galaxies
$d(U-V)/dM_V = -$0.08 \citep{ble92,terlevich01}.
This slope appears compatible with the CMRs presented
in Fig.\ \ref{fig:cmr} \citep[a quantitative assessment of 
this statement is challenging without developing a 
robust description of the entire bimodal 
distribution, which is beyond the scope of this work; see, e.g.,][]{baldry04}, 
and indeed is compatible with the slope 
of $-0.05$ derived by \citet{schweizer92} for E and S0
galaxies in field environments.
We determine the robust biweight mean 
color, with the locally-determined CMR slope
subtracted, of all galaxies with $U - V > 1.0$ (independent of redshift, 
so as to roughly isolate the red sequence).
We illustrate the procedure by showing the CMR-subtracted 
color histogram of galaxies
in the inset panels of Fig.\ \ref{fig:cmr}.  
The biweight mean and sigma of the unbinned sample are denoted by
the dashed and dotted lines respectively.  It becomes clear that the 
peak of the distribution can be determined to within a few hundredths of a
magnitude, which is sufficient for our purposes. 

The value of our fit to the CMR color at $M_V - 5\log_{10}h= -20$ 
is shown as a function of redshift in Fig.\ \ref{fig:cmrint} and 
is tabulated in Table \ref{tab:fits}.  
It is clear that this 
population evolves considerably between $z = 1.1$ and 0.2; 
the galaxies become redder
by $\sim 0.3$ mag.  For completeness, a linear fit to the 
evolution (including the local comparison points) is
$\langle U-V \rangle_{M_V = -20} = 1.40 - 0.31z$, as shown 
by the solid grey line.  Some idea of the systematic uncertainties
is given by a fit to the \combo points only: 
$\langle U-V \rangle_{M_V = -20} = 1.48 - 0.40z$ (the fit is not shown 
for clarity).  The biweight RMS for the red sequence
is given by the error bar.  Overplotted are the 
evolving $U-V$ colors expected for single-age populations
with differing formation redshifts.  We use the 
\peg stellar population synthesis model with subsolar metallicities  
to predict colors at a constant absolute magnitude. 
Galaxies with $M_V - 5 \log_{10} h = -20$ at 
redshift zero would be brighter at redshift one, owing to 
passive evolution of the stellar population.  Therefore, 
we apply a small color correction to the passive evolution models 
to account for the fact that the galaxies with $M_V - 5 \log_{10} h = -20$
at redshift one are from a different, bluer part of the CMR.
The evolution of the model colors are insensitive to the 
choice of stellar initial mass function (IMF), for which we adopt
the \citet{salpeter} parameterization.
We show also the color of the CMR intercept from two 
local samples. The color intercept of morphologically-classified
early-type galaxies in the Abell 754 galaxy cluster is
shown by the naked error bars \citep{mcintosh03}. This 
color intercept is consistent with 
early-type galaxies from a wide range of environments from 
the Nearby Field Galaxy Survey \citep{jansen00,mcintosh03}.
The $U-V$ CMR intercept synthesized from 
SDSS Early Data Release \citep[EDR;][]{edr}
$u$, $g$, and $r$-band data is shown by the diamond with error bars, 
which is offset from zero redshift for clarity.  
The SDSS EDR sample is discussed in the Appendix.

The color evolution of the \combo CMR intercept (the 
$\langle U-V \rangle$ at $M_V -5 \log_{10} h= -20$) is consistent
with the expectations of passive evolution of ancient
stellar populations.  However, 
the \combo points do seem slightly high
compared to the local data at $z \la 0.5$.  Absolute calibration 
of surveys is challenging at even the $\sim 10$\% level, 
especially as all of the data shown in Fig.\ \ref{fig:cmrint}
have been transformed to $U-V$ from other passbands, except for
the Abell 754 data from \citet{mcintosh03}.  
Furthermore, the number of red-sequence galaxies is
low at $z \le 0.4$, increasing the uncertainties considerably
(see Fig.\ \ref{fig:cmr}).
Finally, and perhaps most
importantly, color gradients will affect 
this analysis at some level.
The \combo colors are measured using adaptive-aperture
photometry, and are essentially small-aperture central colors.  
For the most distant galaxies this will approximate
total color.  For the nearest \combo galaxies at $z \la 0.4$, 
we may sample just
their inner, redder parts.  Taking typical color gradients
from \citet{peletier90} as a guide, this could easily lead to 
a $\sim 0.1$ mag
reddening of the CMR intercept at low redshift compared to the 
passive evolution expectation from high redshift.  
The local surveys we show use colors derived
for between 50\% and 100\% of the galaxy light, and this
would also lead to an offset between the lowest-$z$ \combo and local color
intercepts at around the $\sim 0.1$ mag level.
Both of these effects are arguably seen in the data.
Nevertheless, taken as an ensemble, the 
CMR intercepts do appear to redden with time in a way that is consistent
with the expectations of passive evolution. 
Yet, even extending large samples to $z \sim 1$, it is not possible
to constrain meaningfully the formation redshift of the {\it stars}
in early-type galaxies using this type of information,
owing both to the age-metallicity degeneracy, 
and to systematic model uncertainties of 
at least $\sim 10\%$ in the colors of old 
stellar populations \citep{charlot96}.
This is illustrated in Fig.\ \ref{fig:cmrint}, 
where we choose stellar populations of different ages, 
with their metallicity at $M_V -5\log_{10}h= -20$ adjusted 
to keep the average $U-V$ color evolution roughly constant.

\begin{figure*}[tb]
\hspace{1.3cm}
\epsfxsize=15cm
\epsfbox{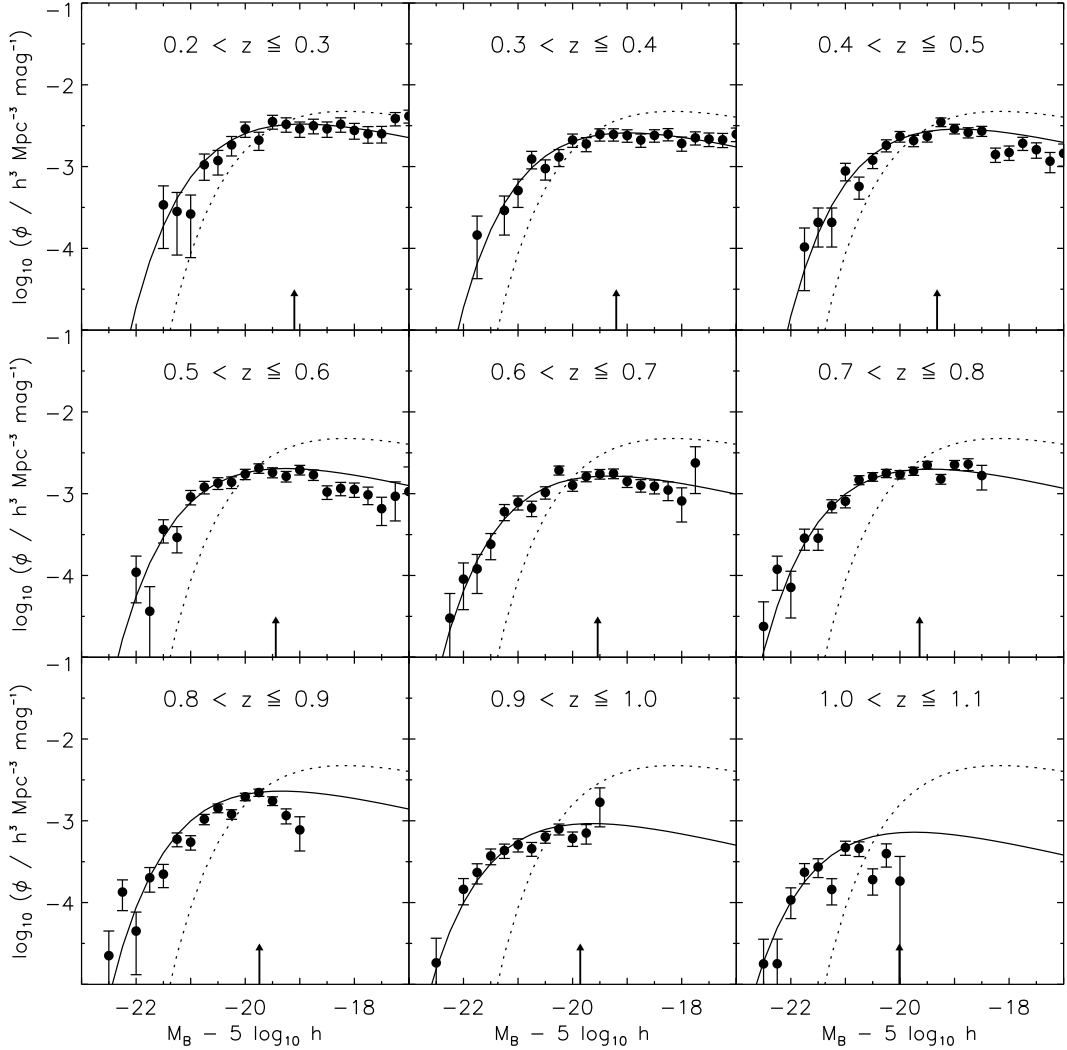}
\vspace{-0.4cm}
\caption{\label{fig:lfevo} Redshift evolution 
in the rest-frame $B$-band luminosity function of red-sequence galaxies.
Black solid lines with data points show the Schechter Function fits and 
$V/V_{\rm max}$ luminosity functions of
\combo red-sequence galaxies in each redshift interval.  The dotted line
shows the Schechter fit to the low-redshift SDSS comparison 
sample (see the Appendix for more details).  A single
faint-end slope $\alpha = -0.6$ is derived
by a single STY fit to the \combo data at all redshifts.
When deriving the $\phi^*$ estimate, the three least luminous
bins are ignored.
The arrow shows the magnitude cutoff used for the 
analysis of the luminous red-sequence galaxy population in \S \ref{sec:origin},
assuming $z_f = 3$. 
 }
\end{figure*}

It is worth briefly discussing the scatter 
in the CMR of early-type galaxies.  In the local
Universe, the scatter in the CMR is observed to be small 
\citep[$\la 0.04$ mag;][although Schweizer \& Seitzer 1992 find a scatter
of $\sim 0.1$ mag in a more field galaxy-dominated sample]{ble92,terlevich01}.
This has been interpreted as evidence for a
small age spread in local Universe early-type galaxies, although 
\citet{trager00} argue that age and metallicity effects 
can conspire to produce a small CMR scatter even with 
large, anti-correlated scatter in age and 
metallicity.  We find
rather larger scatter (Fig.\ \ref{fig:cmr}), mostly because our photometric
redshift errors of $\Delta z \sim 0.03$ translate into errors in
the derived rest-frame colors of $\Delta (U-V) \sim 0.1$\,mag. 
In addition, the red sequence at $z \la 0.4$ is 
poorly defined, leading to larger uncertainties.
Owing to these, and other systematic uncertainties (such as 
calibration and mismatches between real galaxy spectra and the 
template spectra used to derive the $k$-correction), we show 
the full $\pm 1 \sigma$ scatter in Fig.\ \ref{fig:cmrint}. 
In any case, these sources of uncertainty do not significantly 
affect our analysis because the color selection cut of  
$\Delta (U-V) = 0.25$\,mag that we choose later in 
\S \ref{sec:lf} is larger than the scatter at all redshifts ($\la 0.2$ mag).

\section{The Luminosity Density of Red-Sequence Galaxies} \label{sec:lf}

\begin{table*}
\caption{Red-Sequence Galaxy Rest-frame $B$-band 
Luminosity Function Fits and CMR
Intercept {\label{tab:fits}}}
\begin{center}
\begin{tabular}{lr@{$\pm$}rr@{$\pm$}lr@{$\pm$}rccr@{$\pm$}l}
\hline
\hline
$\langle z \rangle$ & \multicolumn{2}{c}{$M_B^*-5$~log~$h$} &
  \multicolumn{2}{c}{$\phi^* \times 10^{-4}$} &
  \multicolumn{2}{c}{$j_B \times 10^{7} L_{\odot}$} & $f_{\rm obs}$ & $c_{\phi^*,L^*}$ &
 \multicolumn{2}{c}{$(U-V)_{M_V = -20}$} \\
 & \multicolumn{2}{c}{(Vega mag)} & \multicolumn{2}{c}{$(h/$Mpc)$^{-3}$}
& \multicolumn{2}{c}{$(h/$Mpc$^3$)} & & & \multicolumn{2}{c}{} \\
\noalign{\smallskip} \hline \noalign{\smallskip}
0.25 & $-19.94$ & $ 0.22$ & 71 & 16, 20 & $
 8.54$ & 1.88, 2.34 & 1.0 & $-0.297$ & 1.41 & 0.22 \\
0.35 & $-19.98$ & $ 0.19$ & 56 & 23, 12 & $
 6.87$ & 2.81, 1.42 & 1.0 & $-0.291$ & 1.35 & 0.19 \\
0.45 & $-19.92$ & $ 0.17$ & 62 & 9, 10 & $
 7.27$ & 1.10, 1.23 & 0.99 & $-0.300$ & 1.30 & 0.18 \\
0.55 & $-20.25$ & $ 0.17$ & 44 & 2, 6 & $
 6.94$ & 0.38, 1.00 & 0.99 & $-0.251$ & 1.24 & 0.16 \\
0.65 & $-20.34$ & $ 0.17$ & 35 & 7, 5 & $
 6.09$ & 1.14, 0.78 & 0.97 & $-0.315$ & 1.16 & 0.12 \\
0.75 & $-20.44$ & $ 0.16$ & 43 & 3, 5 & $
 8.16$ & 0.54, 0.94 & 0.94 & $-0.456$ & 1.14 & 0.14 \\
0.85 & $-20.33$ & $ 0.14$ & 50 & 11, 5 & $
 8.43$ & 1.91, 0.89 & 0.88 & $-0.675$ & 1.15 & 0.15 \\
0.95 & $-20.64$ & $ 0.17$ & 20 & 5, 2 & $
 4.50$ & 1.11, 0.44 & 0.85 & $-0.767$ & 1.10 & 0.13 \\
1.05 & $-20.74$ & $ 0.22$ & 16 & 3, 1 & $
 3.87$ & 0.64, 0.35 & 0.77 & $-1.019$ & 1.09 & 0.12 \\
\noalign{\smallskip} \hline
\end{tabular}
\\ Note. --- STY fits in 9 redshift intervals where the faint end
slope 
$\alpha = -0.6\pm0.1$ was fit to the entire sample.  Listed are
$M^*$, $\phi^*$ with two estimates of the cosmic variance uncertainty;
from field-to-field variation divided
by $\sqrt{3}$, and following the prescriptions of \citet{somer04}.
The luminosity density $j_B$ is given, again with cosmic
variance uncertainties, and further listed is 
the fraction of the luminosity density
that is constrained by the measurements, and the covariance between 
$\phi^*$ and $M^*$.  Also given is the intercept of 
the CMR at $(U-V)$ color at $M_V = -20$
and the robust biweight RMS scatter.
\end{center}
\end{table*}

Because the stellar populations of red-sequence/early-type 
galaxies may have formed
much earlier than they were assembled into recognizable
early-types \citep{baugh96,vandokkum01}, the 
evolution of the CMR, or of the stellar mass-to-light ratios of 
red galaxies, is not a sensitive diagnostic of
when these galaxies attained their present configuration.
A much more sensitive diagnostic is the density
or luminosity functions of red-sequence galaxies at 
redshifts less than unity.

In \S\S \ref{sec:col} and \ref{sec:cmr}, we have seen 
that it is possible to define the early-type galaxy 
population in an empirical, model-independ\-ent fashion 
by exploiting the bimodality of the galaxy color
distribution out to $z \sim 1$ (Fig.\ \ref{fig:cmr}).
We therefore choose to study galaxies around the red galaxy CMR, 
defining these as the red-sequence galaxy population at any 
given redshift.  More quantitatively, we define
red-sequence galaxies as being redder than 
$\langle U-V \rangle = 1.15 - 0.31z - 0.08 (M_V -5 \log_{10}h+ 20)$
(the dashed line in Fig.\ \ref{fig:cmr}).  This line 
is 0.25 mag bluewards of the linear fit 
to the evolution of the mean color of a $M_V - 5 \log_{10}h = -20$ galaxy
from Fig.\ \ref{fig:cmrint} (the solid grey line)\footnote{
This fit includes the $z \sim 0$ data points also.  A color
cut derived using just the \combo data (which therefore includes
the small bias caused by color gradients) is:
$\langle U-V \rangle = 1.23 - 0.40z - 0.08 (M_V -5 \log_{10}h+ 20)$.
Adoption of this color cut in what follows does not change the 
results to within the errors.}.  
We choose this definition merely
to have a criterion that is a continuous function of 
redshift.  Using a discontinuous color-cut which is 0.25 mag bluewards of
the empirical CMR zero-point defined within each $\Delta z = 0.1$ bin
yields results that agree within the errors.
Our color-cut philosophy is similar to that of 
\citet{bo}, except that we choose
to keep the red galaxies rather than the blue ones.
The advantages of this definition of 
early-type galaxies are many. Firstly, it is straightforward 
to match this selection criterion in theoretical models.  For example, even if 
the model CMR is curved, one can simply go $\sim$0.25 mag redwards of
the curved locus in $U - V$ space and choose galaxies redder than 
that curved locus (although as long as the `gap' between
the red and blue peaks is sparsely populated, like in the observational
data, the exact choice of color cut makes little difference
as long as the cut lies in the `gap').  
Secondly, it automatically accounts for evolution
in a model-free way by fitting the CMR ridge-line.  Thirdly,
this definition has a well-defined physical meaning (galaxies that 
have colors consistent with old stellar populations at a given 
redshift).  A final advantage is this definition's insensitivity to 
differences in photometric zero points.  For example, the local 
SDSS comparison sample and \combo sample may have offset $U-V$
color scales (Fig.\ \ref{fig:cmrint}); yet, the $B$-band luminosity density
of galaxies on the CMR is insensitive to this small color
offset, meaning that the luminosity density of the 
local SDSS sample can be meaningfully compared with 
the rest-frame $B$-band luminosity density of the \combo dataset.
We note that this definition contrasts with W03, who adopted
spectral type definitions based on present-day spectral energy distribution
types. With this new approach, we are taking into account in 
a natural and empirical way the effects of evolution in red sequence
galaxy color.

Defining galaxy types in this way, we now explore the 
evolution of the luminosity function of 4690 $0.2 \le z \le 1.1$ 
\combo red-sequence galaxies in 
Fig.\ \ref{fig:lfevo}.  We estimate the galaxy luminosity function
in different redshift bins following W03 using 
the $V/V_{\rm max}$ formalism \citep[e.g.,][]{felten77}, using
our detailed estimates of galaxy completeness as a function 
of surface brightness, magnitude, redshift and spectral type (see, e.g., 
Figs.\ 6 \& 7 of W03).  Error bars
on the $V/V_{\rm max}$ data points take into account 
Poisson errors only (i.e., not accounting for redshift, 
spectral type and magnitude errors).  \citet{schechter} functions 
were fit to the red-sequence galaxy
sample using the method of \citet[STY hereafter]{sty}.\footnote{In
studies where the luminosity function is derived from a redshift
survey with considerable `thickness', such that galaxies populating
the faint end come from different volumes as the galaxies
populating the bright end, the $V/V_{\rm max}$ can give
misleading results owing to large scale structure.  The STY
estimator does not suffer from this limitation \citep{sty}.  
In this paper, we split the \combo survey data into many
thin shells, therefore the faint
and bright galaxies in each redshift interval come from nearly
identical volumes.  Thus, the $V/V_{\rm max}$ luminosity function 
estimates should also be relatively robust to large scale structure.} 
We choose to fit the entire red-sequence galaxy sample
at all redshifts with one common value of the faint-end slope $\alpha$;
in practice, $\alpha$ is determined mainly by the lower redshift
galaxy samples and is $-0.6\pm0.1$ in this case, in excellent
agreement with local observations \citep[e.g.,][]{madgwick02,bell03}.  
The magnitude at the `knee' of the 
luminosity function ($M^*$) is redshift, but not field-to-field,
dependent, and the density normalization $\phi^*$
is fit separately to each field at each redshift (see Table
\ref{tab:fits} for fit parameters).
The Schechter luminosity function to the local comparison sample is shown
by a dotted line (see the Appendix for more details).

W03 found strong evolution of `early-type galaxies', defined
via redshift-independent spectral energy distribution 
criteria (which did not take into account galaxy evolution).
In contrast, one can see from Fig.\ \ref{fig:lfevo} that the luminosity
function evolution of early-type galaxies defined in this way 
is relatively mild. The characteristic magnitude 
$M^*$ seems to brighten and the number density $\phi^*$ seems to decrease
somewhat with increasing redshift.

\begin{figure*}[tb]
\vspace{-0.5cm}
\hspace{0.5cm}
\epsfbox{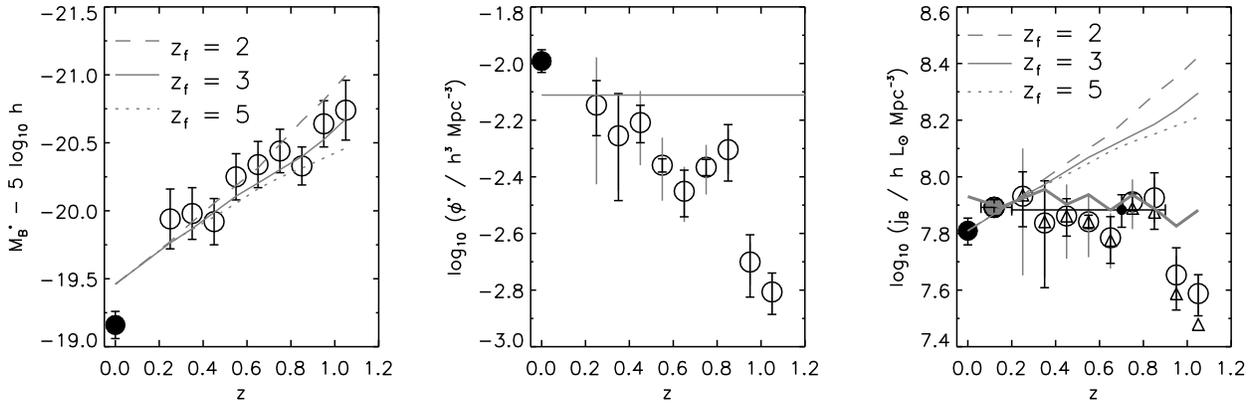}
\vspace{-0.2cm}
\caption{\label{fig:lumdens} The evolution of 
the rest-frame $B$-band luminosity function
of the red-sequence galaxy population from COMBO-17.
{\it Left:} The evolution of $M_B^*$, the `knee' of
the luminosity function.  We show the expectations of 
a passive evolution model as smooth grey lines 
(with formation redshift shown in the 
Figure legend).  The solid point shows the $M_B^*$
of the SDSS low redshift comparison sample, transformed
from $ugr$ data (see the Appendix for more details).
{\it Center:} The evolution of $\phi^*$: again, 
open circles denote \combo data and the solid point 
the local SDSS sample.  The error bars attempt to indicate
uncertainty owing to cosmic variance; the black error bars 
show the observed field-to-field variation divided by $\sqrt{3}$, 
whereas the grey error bars show the predicted cosmic variance 
using the prescription of \citet{somer04}.
A passive evolution model
predicts no evolution in the number density with redshift (grey line).
{\it Right:}  The rest-frame $B$-band luminosity
density per comoving Mpc$^3$ is shown by open circles, with error
bars again denoting the uncertainties from cosmic variance.  
To reduce the cosmic variance uncertainty, we show also the 
averaged $0.2 \le z \le 0.9$ $j_B$ by the solid small black 
circle, which has only a 13\% uncertainty from cosmic variance.
The upwards-pointing triangles denote
a lower limit to the luminosity density, where only the 
observed luminosity density is accounted for (i.e., 
no extrapolation to zero luminosity).
The filled grey point
denotes the luminosity density of spectral-type selected early-type 
galaxies in the 2dFGRS.  
We show also the blind 
prediction of the \citet{cole00} semi-analytic
galaxy formation model as a thick solid grey curve,
where model galaxies were selected using exactly the
same color-cut methodology as used in this paper. }
\vspace{-0.2cm}
\end{figure*}

We quantify this evolution in Fig.\ \ref{fig:lumdens}, where
we show the redshift evolution of the red-sequence galaxy
rest-frame $B$-band luminosity function parameters.
In the left and center panels, we show the evolution of the
magnitude of the `knee' of the luminosity function, $M_B^*$, and
the density normalization of the luminosity function, $\phi^*$.
The solid circle denotes the low-redshift SDSS EDR comparison sample, 
selected using synthesized $U-V$ colors \citep[see the Appendix]{bell03},
and $k$-corrected and evolution-corrected to redshift zero.
The smooth grey lines show
the expectation of passive evolution models, normalized to 
pass through the \combo $M_B^*$ evolution.  The model 
evolution is insensitive to stellar IMF at the 0.1 mag level.
Error bars in $M_B^*$ show the 68\% 
confidence interval.  
The error bars for $\phi^*$ attempt to show the expected 
effect of cosmic variance, and were derived using two methods.
The black error bars show the field-to-field variations
divided by $\sqrt{3}$, whereas the grey error bars show
predictions from the prescription of \citet{somer04}, 
which over this redshift range for our sample of galaxies
depends primarily on the comoving volume.  As Fig.\ \ref{fig:lumdens}
and Table \ref{tab:fits} show, these totally independent error estimates agree
quite well, demonstrating that we have accounted for the effects of cosmic
variance on our results.  It is worth noting that 
$M_B^*$ and $\phi^*$ estimates are tightly coupled to 
give an invariant result for the total luminosity
density $j_B$ \footnote{This is a possible explanation
for the offsets of the SDSS $z=0$ data
point from extrapolations from the \combo data, as the SDSS
control sample may have a slightly different shape, giving 
low $M_B^*$ and high $\phi^*$ estimates.}.
One can immediately see that the $M_B^*$ evolves, like the colors, 
in a way that is consistent with the evolution of an ancient
stellar population.  In contrast, the $\phi^*$ seems to strongly
evolve in the interval $0 < z < 1$, in clear disagreement with 
passively-evolving or pure luminosity evolution models.

Much more robustly measured is the total luminosity density of 
the summed red-sequence galaxy population.
The luminosity density is estimated
by integrating the best-fit Schechter Function
to infinity using $j = \phi^* L^* \Gamma(2+\alpha)$.  
We show the average luminosity density at each redshift from 
red-sequence galaxies (open circles), again
including an error bar showing the uncertainty from cosmic
variance.  We show also a lower limit to 
the luminosity density, derived only from the observed 
luminosity density without extrapolating to
zero luminosity (upwards-pointing triangles).
We show also the SDSS (solid circle) luminosity density, 
and the 2dFGRS summed luminosity density of 
spectral-type selected early-type galaxies \citep{madgwick02} 
as a filled grey circle.  The smooth grey lines show
the expectation of passive evolution models as shown above.  

This figure shows a few key results.  Firstly, it is clear that 
cosmic variance is a considerable source of uncertainty, 
even for a survey with 3 independent fields of $\sim 0.25$ square degree
each.  Cosmic variance is the limiting uncertainty
for our and similar studies of the $0 < z \la 1$ luminosity 
function evolution of red-sequence
or early-type galaxies.  Furthermore, 
\citet{somer04} shows that more than an order of magnitude
increase in comoving volume is required to reduce the cosmic
variance uncertainties by a factor of two, if the areas imaged
are contiguous.  Alternatively, a factor of four more widely-separated
areas could give the same decrease in uncertainty.  
Therefore, it will be necessary to
cover substantially larger areas to similar depths to 
significantly improve on these results.
Secondly, and most importantly, the rest-frame $B$-band luminosity density
in red-sequence galaxies remains more or 
less constant over the interval $0 < z \la 1$.
In an effort to reduce the effects of cosmic variance,
we estimate the average luminosity density in 
red-sequence galaxies in the interval $0.2 \le z \le 0.9$, at the 
risk of over-binning the data:
$j_B (0.2 \le z \le 0.9) = 7.6 \pm 1.0 \times 10^7 L_{\sun} h\,{\rm Mpc}^{-3}$.
The observed luminosity density falls short 
of the passive evolution prediction by 
at least a factor of two by $z \sim 1$, although cosmic
variance and uncertainty in the passive evolution models
preclude a more accurate assessment.  

Of some importance in establishing this result
is the local estimate of red sequence luminosity density.  
This was the primary
motivation for carefully constructing a SDSS comparison sample.
As discussed in \citet{bell03}, this estimate is accurate 
to 10\% in a systematic and random sense, owing largely to the 
excellent completeness properties of SDSS (used for galaxy selection)
and the Two Micron All 
Sky Survey (which is used to normalize the luminosity function 
to the all-sky number density of $10 < K < 13.5$ galaxies).
Furthermore, the $U-V$ colors were synthesized
to allow identical selection criteria to be applied to the low-redshift
and \combo samples, strongly limiting that source of systematic
uncertainty.  Furthermore, the 2dFGRS estimate of \citet{madgwick02}
agrees with our SDSS estimate, lending further credibility 
to the low-redshift comparison data.  Therefore, we believe
that it is safe to conclude that the rest-frame $B$-band
luminosity density of color-selected early-type galaxies
does not significantly evolve in the interval $0 < z \la 1$.  
Furthermore, because a passively-evolving population would
have to fade by a factor of two or three from redshift one to 
the present day in rest-frame $B$-band, this non-evolving $j_B$ 
corresponds to an increase in stellar mass in red galaxies
of a factor of two to three from redshift unity to the 
present day.

\section{Discussion} \label{sec:disc}

\subsection{Comparison with previous work}

The data presented here are based on a larger galaxy sample than 
all previous work in this field combined.  Furthermore,
we adopt an empirically-motivated definition
of `early-type galaxy' which accounts for evolution of
the stars in these galaxies in a natural and model-independent fashion.
Nevertheless, on the whole   
there is pleasing agreement between our conclusions and 
the results of many of the previous studies of this topic.
The essentially passive evolution of the {\it stars} in early-type
galaxies (not the galaxy population as a whole), as estimated from 
the color intercept of 
the CMR, is in excellent agreement with the
colors \citep[e.g.,][]{kodama97,kodama99,vandokkum01a},
line strengths \citep[e.g.,][]{kuntschner00,kelson01,ziegler01},
and stellar M/Ls \citep[e.g.,][]{vandokkum01,treu02} 
of individual early-type galaxies
between $0 \la z \la 1$, as extensively discussed in 
the introduction.

There is also good agreement between our detection of a factor
of two evolution in stellar mass on the red sequence and 
other works in the literature.  Morphologically-selected
surveys \citep{im96,schade99,menanteau99,brinchmann00,im02}
found little evidence for density evolution given the 
small sample sizes ($\la 150$ galaxies) and cosmic variance, and could not 
rule out passive or mild evolution of the type that we see. 
Noting that the most luminous galaxies are red at
all redshifts up to unity (Fig.\ \ref{fig:cmr}), 
our results also agree with a factor of a few decrease in 
the number of very luminous galaxies in $K$-band selected
samples out to $z \sim 1$ \citep[e.g.,][]{drory01}.

Most color-selected surveys also agree with 
our conclusions.  \citet{lilly95} study evolution in the luminosity
function of galaxies with colors redder than a present-day Sbc galaxy.
They find very little evolution, with large uncertainties from 
small number statistics and large-scale structure, 
which is consistent with both our result and passive evolution.
\citet{kauffmann96} re-analyse this result, finding a factor of
three evolution in galaxies with the colors of early-types, 
although \citet{totani98} point out that spectroscopic incompleteness
affects this result. 
\citet{lin99} agree with this picture using a similar technique
with a larger sample of galaxies over a more restricted redshift range.
\citet{pozetti03} used the $K20$ survey
to estimate the evolution of red galaxies out to 
$z \sim 1$; using a total galaxy sample 
of only 546 galaxies from 52 square arcmins, they 
find mild evolution in the number of luminous red galaxies, 
but with error bars of more than a factor of two (not 
accounting for the dominant uncertainties from large scale structure).
\citet{cimatti03} study the properties of extremely red
objects (EROs) in $K20$ (galaxies with the colors of passively
evolving populations), finding 2--4$\times$ less EROs than predicted
from passive models (although with large cosmic variance uncertainty);
furthermore, many of these galaxies have disturbed or 
disk-like morphologies.
\citet{chen03} have used the Las Campanas Infrared Survey 
to estimate the luminosity density 
evolution of red galaxies in the rest frame $R$-band, finding
at most a factor of a few evolution in red galaxy density, although
they were not able to rule out passive evolution with their
sample.  We are consistent with these results, but have been 
able to place much stronger constraints based on our 
factor of at least three improvement in 
sample size and factor of two larger sky coverage, 
compared to the largest of the previous
works.  Importantly, these previous works often neglected the 
uncertainties stemming from large-scale structure; whereas, we
have quantified and accounted for this dominant source of uncertainty.

Finally, we should compare with our own earlier work, W03,
where we explored the same \combo dataset, but used a 
concepually different, non-evolving definition of `early-type galaxies'
(their type 1).  Their non-evolving spectral typing corresponds to
constant rest-frame color cuts at all redshifts; their type
1 corresponds to $U - V \ga 1.35$, and their type 2 corresponds to 
$1.35 \ga U-V \ga 0.95$.  Thus, type 1 galaxies have the colors
of present-day luminous elliptical and lenticular 
galaxies, but are significantly redder than
the reddest galaxies at higher redshift.  Consequently, W03 found
rapid evolution of type 1s, in the sense
that galaxies with the colors of present-day early-types are 
extremely rare in the distant Universe (but primarily because of passive 
evolution in their colors).
Because of these definition differences, the statement by W03
that galaxies redder than a fixed threshold
have a rapidly evolving luminosity density is consistent
with our statement of a roughly non-evolving rest-frame 
$B$-band luminosity density in red-sequence galaxies.

In summary, all of our results are consistent with those presented to
date in the literature, but with much greater precision and 
understanding of the uncertainties owing to our factor of $\ga 3$ larger
sample and sky coverage.

\subsection{The Nature of Red Galaxies}

We now turn to the nature of galaxies in this red population:
are these galaxies red because they contain almost exclusively
old stars, or is there a significant
contribution from more dusty, star-forming galaxies?

At low redshift, the red population is 
overwhelmingly composed of morphologically early-type
galaxies \citep[e.g.,][]{sandage78,ble92,schweizer92,terlevich01,hogg02}.  
For the SDSS survey, \citet{strateva01} find
that $\sim$80\% of the red galaxies are earlier than Sa in 
morphological type.  

At intermediate redshift, most studies have targeted galaxy 
clusters where
most red or non-star-forming galaxies are spheroid
dominated \citep[e.g.,][]{couch98,vandokkum00,vandokkum01}.  
Furthermore,most of the luminosity density
of morphologically-classified early-type galaxies 
in the Hubble Deep Field North at $z \sim 0.9$
is in red-sequence galaxies \citep{kodama99}.
Recently, \citet{bell04} used the GEMS 
\citep[Galaxy Evolution from Morphology and SEDs][]{rix04} 
HST Advanced Camera for Surveys dataset to explore
the morphology of roughly 500 $0.65 \ge z \ge 0.75$ 
red-sequence galaxies, finding that $\ga 80$\% of
the luminosity density in $z \sim 0.7$ red-sequence galaxies 
is from morphologically-classified early-type (E, S0 and some Sa) galaxies.

The nature of red galaxies at higher redshifts $1 \la z \la 2$ is 
somewhat less clear.  A number of studies have explored the nature
of `extremely red objects' (EROs), which have very red colors
consistent with passively-evolving ancient stellar populations
with $1 \la z \la 2$ (but can in principle be a mixed bag
from a variety of different redshifts).  
\citet{cimatti02} use deep spectroscopy to split EROs roughly
50:50 between passively-evolving red-sequence galaxies and dusty
star-forming galaxies in a $K \le 19.2$ sample of 45 EROs.
Complementary analyses are presented by \citet{yan03}, 
\citet{moustakas04}, \citet{gilbank03}, and \citet{cimatti03}, 
who demonstrate that there is roughly an equal three-way 
split between the {\it Hubble Space Telescope} (HST)
morphologies of EROs, such that $\sim 1/3$ are
bulge-dominated, $\sim 1/3$ are disk-dominated, and $\sim 1/3$ have 
disturbed morphologies.  Thus, luminosity densities derived
from red color-selected $1 \la z \la 2$ samples may {\it overestimate}
the stellar mass in early-type galaxies.

This evidence suggests that most
red galaxies to $z \la 1$ are morphologically early-type with dominant
old stellar populations.  In contrast, a significant fraction 
of distant $z \ga 1$ red galaxies may have red colors not because
of a dominant old stellar population, but rather because of dust
(i.e., edge-on disks or dusty starbursts).  Bearing this in mind, 
we caution that our
results should be interpreted only as upper limits on the abundance
of galaxies dominated by ancient stellar populations until the advent 
of more definitive morphological data and/or spectra for sufficient
numbers of galaxies with $0.2 \la z \la 1$.  

\subsection{The Predicted Luminosity Evolution of Hierarchical Model
	Galaxies} \label{sec:hgf}

In \S \ref{sec:lf} we showed that the rest-frame
$B$-band luminosity density of red-sequence
galaxies does not significantly
evolve since $z \sim 1$.  This is at variance
with the expectations of a completely passive evolution model
(where red-sequence galaxies fully form at $z \gg 1$ and simply
age to the present day),
which would predict a factor of two or three decrease
in the rest-frame $B$-band luminosity density in 
red galaxies from redshift one to zero.  It is interesting
to compare our results with the expectations 
of hierarchical models of galaxy formation and evolution, 
to check if these models would predict a roughly non-evolving
$B$-band luminosity density since $z \sim 1$ in 
red galaxies.

There are two main predictions of hierarchical models of galaxy
formation and evolution.  Firstly, early-type 
galaxies in the field should have stellar population of a 
somewhat younger mean stellar age, 
and they should be assembled later
than their clustered counterparts, because
the mergers which give rise to the early-types
happened at earlier times in clusters than in the 
field \citep[where they continue even to the present day; e.g.,][]{baugh96}.
We have chosen not to explore this issue at present, as
constructing local density estimators with a redshift precision of 
$\Delta z \sim 0.03$ is non-trivial.  Other studies may have
detected an offset, in the sense that field
early-types may be younger than cluster early-types, 
using line strength \citep{guzman92,bernardi98} or fundamental
plane constraints \citep{treu02}, although
\citet{hogg04} see no environmental dependence in the CMR
in the SDSS at the 0.01 mag level.  Nevertheless, progenitor bias 
and the age-metallicity degeneracy are 
significant observational worries in studies of this 
kind, and more work is required to firm up these 
interesting results.  A particularly
exciting avenue of research is study of mass-selected
galaxies from gravitational lensing studies.  In these studies,
the sample is selected to have high stellar$+$dark masses,
without regard for their photometric properties (or indeed, for their 
environments).  First results from these surveys have been controversial
but promising.  \citet{kochanek00} and \citet{rusin03} find no evidence
for younger stellar populations in (the same) mass-selected sample, 
compared to galaxy clusters.  \citet{vandeven03} 
re-analyze these data using a more model-dependent
but sensitive approach, accounting for 
progenitor bias, finding younger luminosity-weighted
ages than clustered massive galaxies, in agreement with hierarchical 
model expectations.

The second main prediction of hierarchical models of galaxy
formation is a gradual increase in the stellar mass of the 
early-type/red-sequence
galaxy population until the present day, because of constant build-up 
in stellar mass from galaxy mergers \citep{baugh96,kauffmann96}.  
In Fig. \ref{fig:lumdens}, we show a blind
prediction of evolution of the rest-frame $B$-band luminosity density
of red-sequence galaxies from the \citet{cole00} fiducial 
model using the same color-magnitude
relation methodology adopted in this paper.   
The models are constrained to match the local luminosity function
plus the morphological mix of $L \ga L^*$ galaxies in the local
Universe, therefore the approximate match between the models
and the observations at redshift zero is largely by design.
Observations of luminosity functions 
at non-zero redshifts were not used at any stage
to constrain model parameters, therefore the 
excellent agreement between the model and the data
out to $z \sim 1$ is a significant achievement.  
It is unclear how well other models \citep[e.g.,][]{kauffmann99,somer}
will reproduce these data, owing to the well-known inability of models
of this type to match both the dynamics and luminosity functions
of galaxies simultaneously.  The \citet{cole00} model
tunes to match the luminosity function of galaxies well at the present day,
and is therefore the suitable model with which 
to compare luminosity density 
evolution.  The \citet{somer} and \citet{kauffmann99} models
tune to reproduce galaxy
dynamics at the expense of the galaxy luminosity function, 
and are less suitable for comparisons of this type (but are more
suitable for exploring the evolution of galaxy dynamics).
A more thorough discussion of the stellar mass, dynamical and 
morphological evolution of the red-sequence galaxy population 
is clearly warranted, but is beyond the scope of this paper.
Nevertheless, the current zero-order comparison is rather
encouraging; models that are tuned to reproduce the present-day 
distribution of galaxy luminosities predict no evolution 
in the luminosity density of red-sequence galaxies, 
in agreement with our observations.

\subsection{The Origin of Luminous Red-Sequence Galaxies} \label{sec:origin}

What might these data tell us about the origin of 
luminous red-sequence galaxies?  An intriguing observation 
is that at all redshifts, there are very few blue galaxies
luminous enough to fade into the brightest red-sequence galaxies
(Fig.\ \ref{fig:cmr}).  In this section, we explore what this
may imply.

As a useful tool for this discussion, we construct a
sequence of previously star-forming \peg model galaxies who
have had their star formation suddenly stopped at some point
in the past, to study how galaxies fade and redden with time
(the solid lines with crosses in Fig.\ \ref{fig:cmr}).
We construct a `bright' ($M_V -5 \log_{10}h \sim -21$)
and `faint' ($M_V -5 \log_{10}h \sim -19$) set of 
model galaxies.  These models are not designed to be particularly 
realistic, rather they are discussed only to give a flavor
for how star-forming galaxies may settle onto the red sequence
if their star formation was stopped for some reason.
We discuss the `bright' set of 
model galaxies first, discussing the `faint' set 
afterwards.  The `bright' set of model galaxies
is based on a galaxy with a SFH of the form
$\psi \propto e^{-t/\tau}$, where
$\psi$ is the star formation rate (SFR), 
$t$ is the time elapsed since galaxy formation, 
and $\tau$ is the $e$-folding time of the SFH, 
set in this case to be 4 Gyr (this color would correspond
roughly to a Sb galaxy in the local Universe).  The formation 
redshift is assumed to be $z_f = 2$, and
the metallicity is assumed to be constant
at solar, for simplicity.
The evolution of this
model over cosmic time is shown by the bluest 
cross on the `bright' track in each panel of 
Fig.\ \ref{fig:cmr}.  

The solid line at $M_V -5 \log_{10}h \sim -21$
in each panel of Fig.\ \ref{fig:cmr} connects
this $\tau = 4$\,Gyr model with models which 
have the identical SFH (crosses), except that they have 
had their SF stopped arbitrarily 
after an elapsed time of 90\%, 80\%, 50\% and 10\%
of the age of the galaxy at the redshift interval of 
interest (these models correspond to the second bluest cross to 
the reddest cross, respectively; the bluest cross is the
untruncated model as discussed above).  When the SF is stopped, 
a burst of SF occurs which increases the stellar
mass of the truncated model to the same stellar mass
as the untruncated galaxy.  The `faint' set of models
is analagous, except with 1/3 solar metallicity and 
a constant SFR ($\tau = \infty$), in order to have 
the somewhat bluer colors characteristic of fainter galaxies.

A cursory inspection of these tracks in Fig.\ \ref{fig:cmr} shows
clearly that while faint red-sequence galaxies
may plausibly be faded remnants of 
blue, later-type galaxies which exist at the same epoch,
there are very few blue galaxies bright enough to fade into 
the most luminous red-sequence galaxies
{\it at all redshifts out to $z \sim 1$}.
Thus, luminous red-sequence galaxies cannot have originated from
simple fading of single blue
star-forming galaxies since redshift one.  This 
implies that most luminous red-sequence galaxies 
were either completely formed earlier than $z \sim 1$, or
that these luminous red-sequence galaxies
formed through mergers of galaxies.  

\begin{figure}[tb]
\vspace{-0.5cm}
\hspace{-0.5cm}
\epsfbox{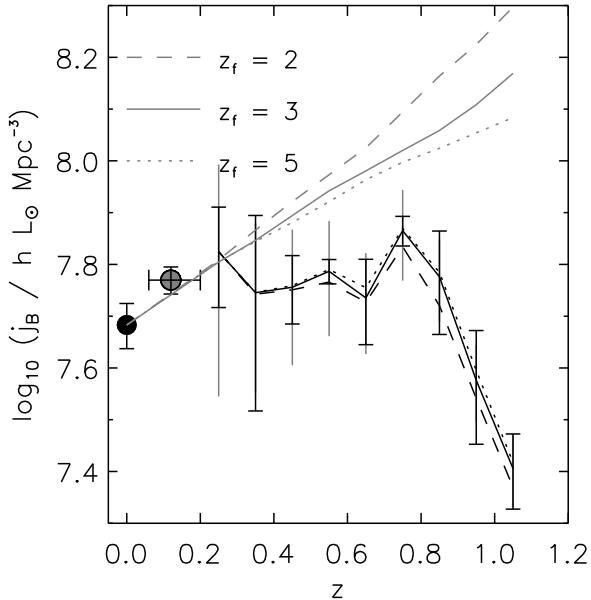}
\vspace{-0.2cm}
\caption{\label{fig:bright} The luminosity density evolution
of luminous red-sequence galaxies.  
The luminosity density of those red-sequence galaxies destined to 
evolve passively into galaxies brighter than 
$M_B - 5\log_{10}h \le -19.1$ at $\langle z \rangle = 0.25$ is
shown by dashed, solid and 
dotted black lines for 
$z_f = 2,3,5$ respectively.  The smooth grey curves show
the expected evolution of a galaxy population that completely
formed at high redshift and simply aged to the present day.  The 
error bars show uncertainties from cosmic variance. 
  } \vspace{-0.2cm}
\end{figure}

We explore these two possibilities by considering
the luminosity density in luminous red-sequence galaxies
at $z \la 1$.  
Let us adopt as a null hypothesis that luminous red-peak 
galaxies form completely at high redshift and simply age
to the present day (the first option).  Then, if we choose to evaluate
the luminosity density in red-sequence galaxies above a passively-fading 
luminosity cut (i.e., a constant stellar mass cut), 
we should observe a luminosity density which fades passively
(like the grey passive evolution tracks in
Figs.\ \ref{fig:lumdens} and \ref{fig:bright}).  
If the total luminosity density in this sub-population does  
not follow a simple fading track,
then a merger origin for at least some of the luminous red
galaxies at $z \la 1$ appears necessary (the second option) to
boost this population of luminous red galaxies.  Recall
that a `bulge$+$fading disk' scenario is not allowed, as
a reservoir of very luminous blue galaxies
(blue because of their disks) would be required to fade
and redden into red-sequence galaxies, in clear contradiction
with the observations (Fig.\ \ref{fig:cmr}).
In order to place the most interesting
constraints, we choose a luminosity cut 
corresponding roughly to the `bright' $M_V -5 \log_{10}h \sim -21$ track
in Fig.\ \ref{fig:cmr} above which there are few blue 
galaxies which can fade onto the red sequence. 

The luminosity density of red-peak galaxies destined to 
passively evolve into galaxies brighter than 
$M_B - 5\log_{10}h \le -19.1$ at $\langle z \rangle = 0.25$ 
(as calculated using the rate of fading of \peg
model stellar populations in conjunction with an assumed
formation redshift) 
is shown in Fig.\ \ref{fig:bright} by dashed, solid and 
dotted black lines for different formation redshifts 
$z_f = 2,3,5$.  
This luminosity cut is equivalent to $M_V - 5\log_{10}h \le -20$,
accounting for the average rest-frame color of red-peak galaxies at that
redshift of $B-V = 0.9$.
We show example cuts as a function of redshift
in Fig.\ \ref{fig:lfevo} as arrows, assuming $z_f = 3$. For reference,
these example cuts are $M_B - 5\log_{10}h \le -$19.1, $-$19.2, $-$19.32,
$-$19.44, $-$19.54, $-$19.64, $-$19.74, $-$19.86, $-$20.01 for 
$\langle z \rangle = $0.25, 0.35, 0.45, 0.55, 0.65, 0.75, 0.85, 0.95, 1.05, 
again for $z_f = 3$.
We overplot large scale structure
errors, as before.  We show also the luminosity density in 
galaxies brighter than $M_B - 5\log_{10}h \le -18.8$ at $z = 0$ 
for SDSS EDR red-peak galaxies as a solid point; this cutoff
corresponds to 
$M_B - 5\log_{10}h \le -19.1$ at $\langle z \rangle = 0.25$, 
passively faded to the present day.  The passive fading 
expectation is shown in the grey lines.  The luminosity densities
are sums of the $V/V_{\rm max}$ points, except for a tiny contribution
to the two highest redshift bins, where we must extrapolate
slightly using the Schechter function fit.

It is clear that the observed luminosity density in {\it luminous}
red peak galaxies evolves relatively little
over the interval $0 \la z \la 1.1$ (similarly to the 
luminosity density evolution in the whole red-sequence
galaxy population).  There is 
perhaps a modest increase out to $z \sim 0.7$, then 
a significant drop at higher redshift.  This weak evolution
does not agree with the expectations of passive evolution models
(where the galaxy population forms completely at high 
redshift and simply ages to the present day).
Thus, the population of luminous red-peak galaxies (those
destined to become red-peak galaxies 
with $M_B - 5\log_{10}h \le -19.1$ by $\langle z \rangle = 0.25$)
grows in stellar mass by roughly a factor of two 
in the interval $0 < z \le 1.1$.  
There is a hint that much of this stellar mass build-up happens
at $z \ga 0.7$, with a modest gradual increase at later times.
Clearly, however, a larger survey with lower uncertainties from 
large scale structure is required to further constrain the epoch
of rapid stellar mass assembly in the luminous red galaxy population.

The significance of the evolution in the {\it luminous} red
galaxy population is that we have chosen a luminosity
cutoff which corresponds to the `bright' truncation model
of Fig.\ \ref{fig:cmr}.  
At all redshifts, it is clear that there are very few single 
blue galaxies bright enough to fade into a red
galaxy with luminosities brighter than this model.  
Yet, this luminous red galaxy population builds up in stellar mass by a factor 
of two in this redshift range.  Red massive
progenitors would have been included in the luminosity
density estimates, and blue massive progenitors 
are observed to be very rare indeed (Fig.\ \ref{fig:cmr}).
Thus, merging between less luminous galaxies is required
to build up these luminous red galaxies 
\citep[see also, e.g.,][for discussion of the increased 
importance of merging in galaxy 
evolution at $z \sim 1$]{lefevre00,patton02,brinchmann00}.  Furthermore,
owing to the rarity of luminous blue galaxies, 
these mergers must have at most a very brief luminous blue phase.
This is possible if
most mergers of massive galaxies are either 
very dusty (obscuring and reddening the stellar populations
until the stellar populations age and redden 
sufficiently to join the red sequence) or are very gas poor to begin with
\citep[e.g., the elliptical-elliptical mergers 
observed by][and explored theoretically by 
Khochfar \& Burkert 2003]{vandokkum01a}.
Exploring these two options in detail is 
far beyond the scope of this work, except to
note that both have clear observational signatures. 
The highly-obscured gaseous mergers should 
be among the brightest far-infrared sources
at any given redshift, and will therefore
be well-constrained by current and planned observations in the 
far-infrared and sub-mm regime \citep[e.g.,][]{flores99}.  
On the other hand, the elliptical-elliptical
mergers, if they build up the galaxy mass by more 
than a factor of two or so, will flatten the 
CMR at the highest luminosities \citep{bower98}.  Disentangling
the contributions from both processes will be a fascinating
prospect for the coming years.

The distribution of galaxy colors also strongly
constrains the `frosting'
model of \citet{trager00}.  In this scenario, many early-types
have residual low-level SF at $z \la 1$.  One could postulate
that the factor of two evolution in stellar mass in red-sequence galaxies
is due to residual SF in the other half of the galaxies,
making them too blue to satisfy our redshift-dependent color cut.  This 
residual SF slows to the present day, giving an apparent evolution 
of a factor of two in stellar mass while avoiding the need for 
wholesale galaxy assembly through galaxy mergers.  However, this
mechanism could not work for at least high-luminosity red
galaxies.  A `frosting' of
SF on top of a massive red-sequence galaxy at $z \sim 1$ would produce an 
even brighter blue galaxy, in clear contradiction with the 
observations.  Indeed, as estimated using the \peg stellar
population model, a color difference of $\Delta (U-V) \sim 
0.25$ (the color cut adopted in our definition of red-sequence galaxies)
corresponds to an ongoing SF rate of only a few percent of 
the average SF rate at earlier times (i.e., the birthrate
$b = {\rm SFR}/\langle {\rm SFR} \rangle < 0.05$).
Thus, the observations strongly limit 
the amount of residual SF in massive red-sequence galaxies all the way from
$z \sim 1$ to the present day to less than a few percent enhancement in stellar
mass over the whole redshift range $0 < z \le 1.1$.  Our observations
say little about less luminous early-types; medium-to-high resolution,
high S/N spectroscopy will allow us to more meaningfully constrain
the SFHs of less luminous red-sequence galaxies.  

\subsection{A Synthesis} \label{sec:syn}

It is interesting to bring together 
some of the different threads in this paper into
an overall qualitative, perhaps in parts naive, picture of galaxy evolution.

Firstly, the color distribution is bimodal.  As discussed above at
the end of \S \ref{sec:origin},
birthrates (the relative present-day SFR compared to the 
average past SFR) for red-sequence galaxies are low $\la 5\%$.  Interesting 
lower limits on the birthrates of blue-peak galaxies can be derived
by assuming that these galaxies are dust-free.  Again using the 
\peg stellar population models, the red end of the blue peak (with 
$U-V \sim 0.7$) corresponds to $b \sim 0.1$, whereas the bluest
galaxies correspond to $b \sim 1$ \citep[using \peg models
or the results of][to transform $U-V$ into $b$]{kennicutt94}
\footnote{It should be noted that the 4th column of their 
Table 1 is mistakenly labelled $U-V$, when it should be
labelled $U-B$.}.  Thus, galaxies
on the blue peak {\it need not be forming stars vigorously at
the present day} \citep[see also][]{brinchmann04}.  
Indeed, modest rates of only 10\% of 
the past average SFR are required to keep a galaxy 
on the blue peak (more, if the effects of dust are important).
Thus, bimodality in the color distribution
of galaxies simply betrays the fact that young stars are
very bright, and that if even a small number of young stars
are present they dominate the optical colors of that galaxy.  
The red peak, in contrast, tells us that there is a substantial population of
non-star forming galaxies, with present day SFRs of 
at most a few percent of their past average SFRs \citep[for example,][find 
that half or more of the stellar mass in the local Universe is in non-star
forming early-type galaxies]{hogg02,kauffmann03,bell03}.  

Secondly, the $B$-band luminosity density
of galaxies in the red sequence does not significantly evolve
in the interval $0 < z \le 1.1$.  This 
implies a build-up of stellar mass on the 
non-star-forming red peak by 
a factor of at least two since $z \sim 1$.  This stellar
mass must come from the blue peak galaxies, 
as we are not permitted to form the stars {\it in situ} 
because of the red colors of the red-sequence population.  

Is the relative number of galaxies in the red and
blue peaks, and in the gap, consistent with this picture?
We explore this by constructing a very simplistic model,
in which some small fraction of
star-forming galaxies in the blue peak are 
`turned off' (by some unspecified mechanism)
and subsequently redden and fade \citep[see, e.g.,][for a model of this type applied to rich galaxy clusters]{kodama01}.  Fading
across the `gap' takes around $\sim 40\%$ 
of the Hubble time (as seen by the crosses
on the fading tracks in Fig.\ \ref{fig:cmr}).
In order to reproduce the roughly factor of two to three
difference in the number of galaxies in the gap compared to 
the blue peak, it is necessary to turn off roughly $5-10\%$ of
the blue galaxies per Gyr, letting them fade across the gap.
This rate of transformation also gives a factor of roughly two
increase in the stellar mass density on the red sequence.
Thus, a very simplistic picture where some small fraction
of the blue population of galaxies 
have their SF stopped every Gyr (by some unspecified mechanism),
subsequently fading into red galaxies, holds water.

Of course, we know that there must be other
physical processes at work.  A truncation-only
model predicts that the characteristic luminosity of
red galaxies is fainter than the blue population.
Yet, the brightest and most
massive present-day galaxies in the Universe are almost all 
red \citep[e.g.,][]{kochanek01,bell03}.  Furthermore,
this situation persists out to $z \sim 1$ (Fig.\ \ref{fig:cmr}).
Another prediction is that
the morphologies of the red and blue peak populations 
should be rather similar, which again is completely
ruled out by observations \citep{strateva01}.  Indeed, 
the hypothesis that {\it luminous} red-sequence galaxies at
$z = 0$ are faded remnants of bulge$+$blue disk systems
at $z \sim 1$ whose disks have faded and reddened can 
be ruled out owing to the paucity of luminous blue galaxies
at $z \sim 1$.  Galaxy mergers are a likely 
process for rectifying these shortcomings. 
Galaxy mergers will increase the characteristic luminosity
of post-merger galaxies through the addition of the pre-existing
stellar populations, in conjunction with any SF associated with
the merger itself.  
Furthermore, the violent relaxation associated with galaxy mergers 
leads to more spheroidal-dominated morphologies
\citep[e.g.,][]{els,toomre,barnes96}.  Indeed, mergers and galaxy interactions
may even provide a mechanism for the truncation of SF, through the 
consumption of the gas in an induced starburst 
\citep[e.g.,][]{barnes96,barton00}.  The only constraint
which we can place on the role and frequency of these mergers
with these data is that the mergers responsible for 
the most luminous red galaxies cannot have a long blue phase,
owing perhaps to dust or a lack of gaseous content in 
the progenitors.

This qualitative picture agrees well with many of the features
of current models of galaxy formation in a hierarchical CDM Universe
\citep[e.g.,][]{white91,kauffmann96,somer,cole00}.  Furthermore,
it makes a number of predictions that are testable in the 
near future.  For example, the merger origin of the most luminous
red galaxies can be tested by both deep optical HST imaging
over wide fields and by far-infrared and sub-mm deep imaging
to constrain the frequency of mergers between gas-poor early types
and more gaseous later types, respectively \citep[building 
on the studies of, e.g.,][]{flores99,lefevre00}.  Additionally, 
one can test the viability of the truncation model, and constrain
truncation mechanisms, through a combination of deep HST imaging
(for galaxy morphologies as a function of color and magnitude)
and deep spectroscopy (to check for signatures of recently-truncated SF, such
as enhanced Balmer absorption lines), 
extending the cluster-based work of, e.g., \citet{poggianti99} to the more
general field environment.  Furthermore, 
one can seek to explore these different galaxy populations
in different local environments, to seek the signature
of more rapid environmentally-induced transformation in 
denser environments, extending to higher redshift
the important work of, e.g., \citet{lewis02},
\citet{gomez03}, and \citet{balogh04}.

\section{Conclusions} \label{sec:conc}

The evolution of red-sequence/early-type galaxies directly reflects
the importance and number of major galaxy mergers, and
therefore strongly constrains hierarchical models of galaxy
formation and evolution.  We have explored
the rest-frame colors and luminosities of $\sim 25000$ $R \la 24$
galaxies from 0.78 square degrees 
from the \combo survey, encompassing 
a cosmologically-representative total volume of 
$\sim 10^6 h^{-3}$\,Mpc$^3$.  We find that the distribution of 
galaxy colors is bimodal at all redshifts out to $z \sim 1$.

The blue star-forming peak has colors that become redder
towards the present day at a given magnitude, indicating
changes in characteristic stellar age, metallicity or dust content
with epoch.  Furthermore, there are
many more luminous blue galaxies at $z \ga 0.5$ than 
there are at the present day \citep[this is discussed 
in much more detail by][]{wolf03}.  

The red non-star-forming peak forms a scattered but well-defined
color-magnitude relation (CMR) at $0.2 < z \le 1.1$, with
luminous galaxies being redder than their fainter counterparts.
The color of red galaxies at a given rest-frame magnitude
becomes bluer with increasing redshift.  The quantitative
size of this change is consistent with passive aging of 
ancient stellar populations to the present day.  

Using this empirically-motivated
definition of early-type gal\-ax\-ies, we estimate the rest-frame $B$-band 
luminosity function and luminosity density of red-sequence galaxies 
in the interval $0 < z \le 1.1$.  
There may be some contribution 
from dusty star-forming galaxies at the highest redshifts,
making our results upper limits.  Nevertheless, we find mild evolution
of the rest-frame $B$-band luminosity density between 
$0 < z \le 1.1$.  Ancient stellar populations would
fade by a factor of two to three in this time interval; 
therefore, this mild evolution betrays an increase in the stellar mass
on the red sequence since $z \sim 1$ by at least a factor of two.  
This evolution is consistent qualitatively and quantitatively
with the evolution expected from the hierarchical build-up of
stellar mass via galaxy mergers in a $\Lambda$CDM Universe.
The largest source of error
is large-scale structure, implying that considerably
larger surveys are necessary to further refine this result.

Finally, we explore the evolution of the red, blue and 
gap population on the color--magnitude plane. We find
that a scenario in which SF is stopped (for an
as-yet-unspecified reason) in 5--10\% of the blue
galaxy population per Gyr reproduces the relative numbers of 
red, blue and gap galaxies.  We argue also, based
on the luminosity functions and morphologies of red
galaxies in the local and distant Universe, that 
galaxy merging plays an important role.  In particular,
at least some of the the most luminous red galaxies must be formed 
in galaxy mergers which are either very dusty or 
take place between gas-poor progenitors.  Disentangling
further the processes which drive the evolution of the
red galaxy population at $z \la 1$ will indeed present a fascinating
observational challenge over the next few years.

\acknowledgements

E.\ F.\ B.\ was supported by the European
Community's Human Potential Program under contract
HPRN-CT-2002-00316, SISCO.
C.\ W.\ was supported by the PPARC rolling grant in Observational
Cosmology at University of Oxford and by the DFG--SFB 439.
D.\ H.\ M.\ acknowledges support from the National Aeronautics
and Space Administration (NASA) under LTSA Grant NAG5-13102
issued through the Office of Space Science.
We wish to thank Carlton Baugh for providing blind
predictions of the \citet{cole00} {\sc galform}
semi-analytic model, and thank David Anderson, Marco Barden, Stephane Carlot,
Shaun Cole, David Hogg, Ulrich Hopp, Guinevere Kauffmann, Cedric Lacey, 
Alvio Renzini, Tom Shanks, and Scott Trager 
for useful discussions and suggestions.  Jim Peebles
is thanked for his comments on an earlier version 
of the manuscript.
This publication makes use of the {\it Sloan Digital
Sky Survey} (SDSS).
Funding for the creation and distribution of the SDSS 
Archive has been provided by the Alfred P.\ Sloan Foundation, the 
Participating Institutions, the National Aeronautics and Space Administration,
the National Science Foundation, the US Department of Energy, 
the Japanese Monbukagakusho, and the Max Planck Society. The SDSS
Web site is \texttt{http://www.sdss.org/}.  The SDSS Participating
Institutions are the University of Chicago, Fermilab, the Institute 
for Advanced Study, the Japan Participation Group, the Johns Hopkins
University, the Max Planck Institut f\"ur Astronomie, the Max
Planck Institut f\"ur Astrophysik, New Mexico State University, 
Princeton University, the United States Naval Observatory, and 
the University of Washington. 
This publication made use of NASA's Astrophysics Data System 
Bibliographic Services.

\appendix
\section{Red-Sequence Galaxies in the Local Universe} \label{app}

\begin{figure*}[tb]
\hspace{1.3cm}
\epsfxsize=15cm
\epsfbox{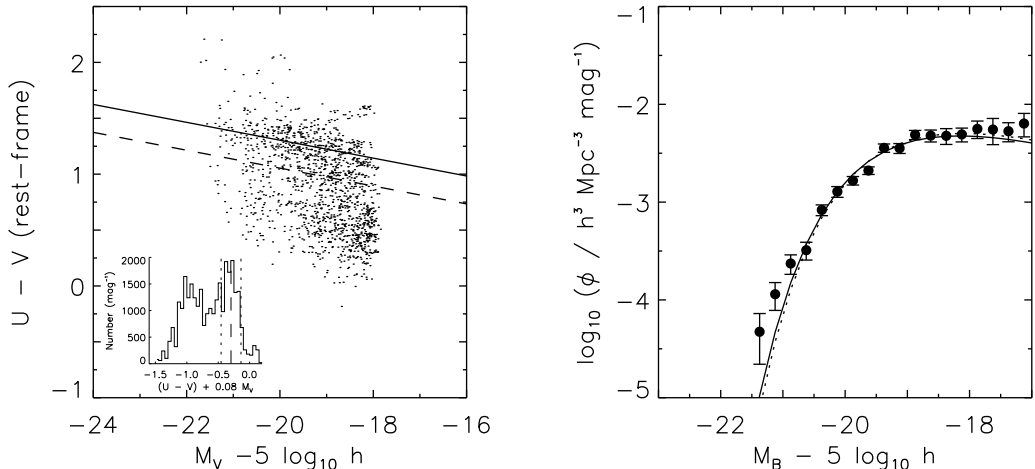}
\vspace{-0.4cm}
\caption{\label{fig:local}   
Rest-frame synthesized $U-V$ colors and $B$-band luminosity
function of SDSS EDR galaxies.  The left-hand panel
shows the synthesized $U-V$ vs.\ $M_V$ color-magnitude 
diagram for a volume-weighted sample of $14.5 \le r \le 16.5$ SDSS galaxies.
We add random offsets of $\sigma_{(U-V)} = 0.01$ and $\sigma_V = 0.1$
for clarity, thus giving clumps of data points for galaxies 
which are included multiple times owing to their small
accessible volume in SDSS.
The solid and dotted lines show the best-fit CMR and the 
Butcher-Oemler-style cutoff adopted in this paper, assuming
a slope of $-0.08$ \citep[following][]{ble92,terlevich01}.
The inset panel shows the color distribution of these galaxies, with 
the slope of the CMR taken out.  
The right-hand panel shows the rest-frame $B$-band
luminosity function of the red-sequence galaxies, assuming a 0.3 mag
magnitude error caused by photometric redshift
uncertainty.  The solid circles with error bars show the $V/V_{\rm max}$
estimates, and the solid line a Schechter fit to these estimates.
The dotted line shows the Schechter fit to the $B$-band luminosity
function without the additional 0.3 mag scatter.
 }
\end{figure*}

Because of its homogeneity and color information, the
SDSS Early Data Release \citep[EDR;][]{edr}
offers an ideal comparison sample to \combo at redshift zero.  
In order to compare as fairly as possible with COMBO-17, we 
choose to synthesize $U$, $B$, and $V$ magnitudes from
SDSS $ugr$ data, using the $UBV$ bandpasses that were used to 
construct COMBO-17 rest-frame magnitudes.  
We adopt the transformations:
\begin{eqnarray}
U - V = -0.713 + 0.826 (u-r); \sigma \sim 0.023, \\
g - B = -0.155 - 0.370 (g-r); \sigma \sim 0.012, \\
g - V = -0.010 + 0.609 (g-r); \sigma \sim 0.006,
\end{eqnarray}
which were derived from the single stellar populations from the 
\peg stellar population synthesis model, and
are valid for stellar populations between 1/200 and 5$\times$ solar
metallicity, and ages between 50\,Myr and 20\,Gyr.  
These transformations are dominated by the systematic zero point
uncertainties (in absolute terms) of the SDSS and Johnson 
photometric system, and by limitations of the SDSS 
Petrosian magnitude. These uncertainties are likely
$\la 0.1$ mag in terms of color, although absolute
calibration of $u$ or $U$-band data is always
challenging.  SDSS EDR galaxies were chosen
to have measured redshifts and a reddening-corrected
Petrosian magnitude in $r$ between $14.5 \le r \le 16.5$,
yielding a total sample of 3592 galaxies.
The Petrosian magnitude of concentrated $c_r \ge 2.6$ (i.e.\ early-type)
galaxies is corrected by $-0.10$ mag to crudely account for 
the well-known shortfall of Petrosian magnitudes for
centrally-concentrated galaxies \citep[e.g.,][]{blanton03a}.

We show the distribution of synthesized $U-V$ color against
$M_V$ for a volume-limited SDSS subsample in the 
left-hand panel of Fig.\ \ref{fig:local}.
Galaxies are weighted by the inverse of the volume within 
which they are observable ($1/V_{\rm max}$), and are selected 
to have a uniform distribution in terms of $1/V_{\rm max}$.
Since SDSS is an apparent magnitude-limited survey, only a small fraction
of the bright, distant galaxies are selected for inclusion
on this plot, whereas the fainter galaxies which are visible
over a much smaller volume are selected multiple times.
We choose to construct a sample of 2000 galaxies with 
$M_V - 5 \log_{10} h \le -18$ (to roughly match the magnitude cutoff
for galaxies at $z \sim 0.6$), and we add random color
and magnitude offsets with $\sigma_{(U-V)} = 0.01$ and 
$\sigma_V = 0.1$ to allow the reader to see multiply-included
galaxies explicitly.  This SDSS control sample shows a
CMR with scatter 0.16 mag;
this large scatter is expected, as the total $u-r$ colors we used
have typical uncertainties $\sim 0.15$ mag, after observational 
error and evolution and $K$-correction errors are accounted for.
The intercept of the CMR is consistent with the intercept
of the Abell 754 galaxy cluster \citep{mcintosh03} and the
Nearby Field Galaxy Survey \citep{jansen00} morphologically-classified
early-type galaxy CMR to within 0.1 mag. This kind of offset is 
not unexpected given the difficulty of calibrating absolute photometry
to better than 10\%.  

Following \citet{bell03}, we construct a rest-frame $B$-band
luminosity function of red-sequence galaxies (redder than the dashed
line in Fig.\ \ref{fig:local}) using the $V/V_{\rm max}$ formalism.  
Galaxy selection for inclusion in this luminosity function
is unbiased by any zero-point uncertainty in calibration, as this
criterion is applied relative to the ridge-line of the CMR, which 
is measured using the same data.  In order
to constrain the effects of magnitude error (from photometric
redshift error) on the luminosity function \citep[as discussed
by][]{chen03},
we convolve our $B$-band luminosities with a random 0.3 mag error, 
which simulates a $\delta z \sim 0.02$ at $z \sim 0.25$ (our lowest
redshift bin).  We also constrain the faint-end slope $\alpha = -0.6$, 
to match with the \combo luminosity function determinations.
The Schechter function fit parameters are $\phi^* = 10.2\pm0.9 \times
10^{-3} h^3\,{\rm Mpc}^{-3}\,{\rm mag}^{-1}$, 
$M^*_B -5\log_{10}h= -19.16\pm0.10$, $\alpha = -0.6$,
and $j_B = 6.4\pm0.2\times 10^7 L_{\sun} h\,{\rm Mpc}^{-3}$ (random
errors only).
The luminosity function 
calculated in this way is shown by the solid
black circles and solid line 
in the right-hand panel of Fig.\ \ref{fig:local}.  For reference, 
a Schechter function fit to the unconvolved (i.e.\ real) $B$-band
synthesized luminosity function is shown by the dotted line.
The $M^*$ of this fit is $\sim 0.05$ mag fainter than the convolved
fit (with corresponding increase of $\phi^*$ to compensate); 
this modest offset is the maximum bias that we expect
in the Schechter fit parameters because of our use of 
17-passband derived photometric redshifts.  This is in 
agreement with \citet{chen03}, who show that
we would expect only a modest bias towards brighter $M^*$ 
because of our relatively small photometric redshift uncertainty.
The uncertainty in luminosity density is roughly 10\% 
in a systematic sense; this error is dominated by
uncertainties in absolute magnitude calibration and
in correcting the luminosity function normalization
to the observed all-sky $|b| \ge 30\arcdeg$ density
of $10 < K < 13.5$ galaxies.

For more discussion of the selection, photometric
parameters, $V/V_{\rm max}$ luminosity function estimation,
and uncertainties, see \citet{bell03}.


\begin{thebibliography}{}

\bibitem[Aragon-Salamanca, Baugh, \& Kauffmann(1998)]{aragon98}
	Aragon-Salamanca, A., Baugh, C.\ M., \& Kauffmann, G. 1998, 
	\mnras, 297, 427

\bibitem[Baade et al.(1998)]{baade98}
	Baade, D., et al. 1998, The Messenger, 93, 13

\bibitem[Baade et al.(1999)]{baade99}
	Baade, D., et al. 1999, The Messenger, 95, 15

\bibitem[Baldry et al.(2004)]{baldry04}
	Baldry, I.\ K., Glazebrook, K., Brinkmann, J., Ivezic, Z., 
	Lupton, R.\ H., Nichol, R.\ C., Szalay, A.\ C. 2004, \apj,
	in press (astro-ph/0309710)

\bibitem[Balogh et al.(2004)]{balogh04}
	Balogh, M., et al. 2004, submitted to \mnras { }(astro-ph/0311379)

\bibitem[Barnes(1992)]{barnes92}
	Barnes, J.\ E. 1992, \apj, 393, 484

\bibitem[Barnes \& Hernquist(1996)]{barnes96}
	Barnes, J.\ E., \& Hernquist, L. 1996, \apj, 471, 115

\bibitem[Barton, Geller, \& Kenyon(2000)]{barton00}
	Barton, E.\ J., Geller, M.\ J., \& Kenyon, S.\ J. 2000, \apj, 530, 660

\bibitem[Baugh, Cole, \& Frenk(1996)]{baugh96}
	Baugh, C.\ M., Cole, S., \& Frenk, C.\ S. 1996, \mnras, 283, 1361

\bibitem[Bell \& de Jong(2000)]{bdj}
	Bell, E.\ F., \& de Jong, R.\ S. 2000, \mnras, 312, 497 

\bibitem[Bell et al.(2003)]{bell03}
	Bell, E.\ F., McIntosh, D.\ H., Katz, N., \& Weinberg, M.\ D. 2003, 
	\apjs, 149, 289

\bibitem[Bell et al.(2004)]{bell04}
	Bell, E.\ F., et al. 2004, \apj, 600, L11

\bibitem[Bernardi et al.(1998)]{bernardi98}
	Bernardi, M., et al. 1998, \apj, 508, L143

\bibitem[Bernardi et al.(2003)]{bernardi03}
	Bernardi, M., et al. 2003, \aj, 125, 1882

\bibitem[Bertin \& Arnouts(1996)]{sex}
	Bertin, E., \& Arnouts, S. 1996, \aaps, 117, 393


\bibitem[Blanton et al.(2003a)]{colblanton}
	Blanton, M.\ R., et al. 2003a, \apj, 594, 186

\bibitem[Blanton et al.(2003b)]{blanton03a}
	Blanton, M.\ R., et al. 2003b, \apj, 592, 819

\bibitem[Bower, Kodama, \& Terlevich(1998)]{bower98}
	Bower, R.\ G., Kodama, T., \& Terlevich, A. 1998, \mnras, 299, 1193

\bibitem[Bower, Lucey, \& Ellis(1992)]{ble92}
	Bower, R.\ G., Lucey, J.\ R., \& Ellis, R.\ S. 1992, \mnras, 254, 601

\bibitem[Brinchmann \& Ellis(2000)]{brinchmann00}
	Brinchmann, J., \& Ellis, R. S. 2000, \apj, 536, L77

\bibitem[Brinchmann et al.(2004)]{brinchmann04}
	Brinchmann, J., Charlot, S., White, S.\ D.\ M., Tremonti, C., 
	Kauffmann, G., Heckman, T., \& Brinkmann, J. 2004, submitted
	to \mnras { }(astro-ph/0311060)

\bibitem[Butcher \& Oemler(1984)]{bo}
	Butcher, H., \& Oemler, Jr., A. 1984, \apj, 285, 426

\bibitem[Calzetti et al.(2000)]{calzetti00}
	Calzetti, D., Armus, L., Bohlin, R.\ C., Kinney, A.\ L., 
	Koornneef, J., \& Storchi-Bergmann, T. 2000, \apj, 533, 682

\bibitem[Charlot, Worthey, \& Bressan(1996)]{charlot96}
	Charlot, S., Worthey, G., \& Bressan, A. 1996, \apj, 457, 625

\bibitem[Chen et al.(2003)]{chen03}
	Chen, H.-W., et al. 2003, \apj, 586, 745

\bibitem[Cimatti et al.(2002)]{cimatti02}
	Cimatti, A., et al. 2002, \aap, 381, L68

\bibitem[Cimatti et al.(2003)]{cimatti03}
	Cimatti, A., et al. 2003, \aap, 412, L1

\bibitem[Cole et al.(2000)]{cole00}
	Cole, S., Lacey, C., Baugh, C.\ M., \& Frenk, C.\ S. 2000, 
	\mnras, 319, 168

\bibitem[Couch et al.(1998)]{couch98}
	Couch, W.\ J., Barger, A.\ J., Smail, I., Ellis, R.\ S., 
	\& Sharples, R.\ M. 1998, \apj, 497, 188

\bibitem[Cowie et al.(1996)]{cowie96}
	Cowie, L.\ L., Songaila, A., Hu, E.\ M., Cohen, J.\ G. 1996, \aj, 
	112, 839

\bibitem[Daddi et al.(2002)]{daddi02}
	Daddi, E., et al. 2002, \aap, 384, L1

\bibitem[Dressler(1980)]{dressler80}
	Dressler, A. 1980, \apj, 236, 351

\bibitem[Dressler et al.(1997)]{dressler97}
	Dressler, A., et al. 1997, \apj, 490, 577

\bibitem[Drory et al.(2001)]{drory01}
	Drory, N., Bender, R., Snigula, J., Feulner, G., Hopp, U., 
	Maraston, C., Hill, G.\ J., \& Mendes de Oliveira, C. 2001, \apj, 562,
	L111 

\bibitem[Efstathiou et al.(2002)]{efstathiou02}
	Efstathiou, G., et al. 2002, \mnras, 330, L29

\bibitem[Eggen, Lynden-Bell, \& Sandage(1962)]{els}
	Eggen, O.\ J., Lynden-Bell, D., \& Sandage, A.\ R. 1962, \apj, 136, 748

\bibitem[Felten(1977)]{felten77}
	Felten, J.\ E. 1977, \aj, 82, 861

\bibitem[Fioc \& Rocca-Volmerange(1997)]{fioc97}
	Fioc, M., \& Rocca-Volmerange, B. 1997, \aap, 326, 950

\bibitem[Firth et al.(2002)]{firth02}
	Firth, A.\ E., et al. 2002, \mnras, 332, 617

\bibitem[Flores et al.(1999)]{flores99}
	Flores, H., et al. 1999, \apj, 517, 148

\bibitem[Freedman et al.(2001)]{freedman01}
	Freedman, W.\ L., et al. 2001, \apj, 553, 47

\bibitem[Gilbank et al.(2003)]{gilbank03}
	Gilbank, D.\ G., Smail, I., Ivison, R.\ J., Packham, C. 2003, 
	\mnras, 346, 1125

\bibitem[G\'omez et al.(2003)]{gomez03}
	G\'omez, P.\ L., et al. 2003, \apj, 584, 210

\bibitem[Guzman et al.(1992)]{guzman92}
	Guzman, R., Lucey, J.\ R., Carter, D., Terlevich, R.\ J. 
	1992, \mnras, 257, 187

\bibitem[Hogg et al.(2002)]{hogg02}
	Hogg, D.\ W., et al. 2002, \aj, 124, 646

\bibitem[Hogg et al.(2003)]{hogg03}
	Hogg, D.\ W., et al. 2003, \apj, 585, L5

\bibitem[Hogg et al.(2004)]{hogg04}
	Hogg, D.\ W., et al. 2004, submitted to 
	\apj { }letters (astro-ph/0307336)

\bibitem[Im et al.(1996)]{im96}
	Im, M., Griffiths, R.\ E., Ratnatunga, K.\ U., \& Sarajedini, V.\
	L. 1996, \apj, 461, L79

\bibitem[Im et al.(2002)]{im02}
	Im, M., et al. 2002, \apj, 571, 1361

\bibitem[Jansen et al.(2000)]{jansen00}
	Jansen, R., Franx, M., Fabricant, D., \& Caldwell, N. 2000, \apjs, 
	126, 271 

\bibitem[Kauffmann \& Charlot(1998)]{kauffmann98}
	Kauffmann, G., \& Charlot, S. 1998, \mnras, 297, L23

\bibitem[Kauffmann et al.(1996)]{kauffmann96}
	Kauffmann, G., Charlot, S., \& White, S.\ D.\ M. 1996, \mnras, 
	283, L117

\bibitem[Kauffmann et al.(1999)]{kauffmann99}
	Kauffmann, G., Colberg, J.\ M., Diaferio, A., \& White, S.\ D.\ M. 
	1999, \mnras, 303, 188

\bibitem[Kauffmann et al.(2003)]{kauffmann03}
	Kauffmann, G., et al. 2003, \mnras, 341, 54

\bibitem[Kelson et al.(2001)]{kelson01}
	Kelson, D.\ D., Illingworth, G.\ D., Franx, M., \& van Dokkum,
	P.\ G. 2001, \apj, 552, L17

\bibitem[Kennicutt, Tamblyn, \& Congdon(1994)]{kennicutt94}
	Kennicutt Jr., R.\ C., Tamblyn, P., Congdon, C.\ E. 1994,
	\apj, 435, 22

\bibitem[Khochfar \& Burkert(2003)]{khochfar03}
	Khochfar, S., \& Burkert, A. 2003, \apj, 597, 117L

\bibitem[Kinney et al.(1996)]{kinney96}
	Kinney, A.\ L., Calzetti, D., Bohlin, R.\ C., McQuade, K., 
	Storchi-Bergmann, T., \& Schmitt, H.\ R. 1996, \apj, 467, 38

\bibitem[Kochanek et al.(2000)]{kochanek00}
	Kochanek, C.\ S., et al. 2000, \apj, 543, 131

\bibitem[Kochanek et al.(2001)]{kochanek01}
        Kochanek, C.\ S., et al. 2001, \apj, 560, 566

\bibitem[Kodama \& Arimoto(1997)]{kodama97}
	Kodama, T., \& Arimoto, N. 1997, \aap, 320, 41

\bibitem[Kodama, Bower, \& Bell(1999)]{kodama99}
	Kodama, T., Bower, R.\ G., \& Bell, E.\ F. 1999, \mnras, 306, 561

\bibitem[Kodama \& Bower(2001)]{kodama01}
	Kodama, T., \& Bower, R.\ G. 2001, \mnras, 321, 18

\bibitem[Kron(1980)]{kron}
	Kron, R.\ G. 1980, \apjs, 43, 305

\bibitem[Kuntschner(2000)]{kuntschner00}
	Kuntschner, H. 2000, \mnras, 315, 184

\bibitem[Larson(1975)]{larson75}
	Larson, R.\ B. 1975, \mnras, 173, 671

\bibitem[Le F\`evre et al.(2000)]{lefevre00}
	Le F\`evre, O., et al. 2000, \mnras, 311, L565

\bibitem[Lewis et al.(2002)]{lewis02}
	Lewis, I., et al. 2002, \mnras, 334, 673

\bibitem[Lilly et al.(1995)]{lilly95}
	Lilly, S.\ J., Tresse, L., Hammer, F., Crampton, D., \& 
	Le F\'evre, O. 1995, \apj, 455, 108

\bibitem[Lin et al.(1999)]{lin99}
	Lin, H., et al. 1999, \apj, 518, 533

\bibitem[Madgwick et al.(2002)]{madgwick02}
	Madgwick, D.\ S., et al. 2002, \mnras, 333, 133

\bibitem[McIntosh, Rix, \& Caldwell(2004)]{mcintosh03}
	McIntosh, D.\ H., Rix, H.-W., \& Caldwell, N., 2004, submitted
	to \apj (astro-ph/0212427)

\bibitem[Menanteau et al.(1999)]{menanteau99}
	Menanteau, F., Ellis, R.\ S., Abraham, R.\ G., Barger, A.\ J.,
	Cowie, L.\ L. 1999, \mnras, 309, 208

\bibitem[Moustakas \& Somerville(2002)]{moustakas02}
	Moustakas, L.\ A., \& Somerville, R.\ S. 2002, \apj, 577, 1

\bibitem[Moustakas et al.(2004)]{moustakas04}
        Moustakas, L.\ A., et al. 2004, ApJL, 600, L131

\bibitem[Nulsen \& Fabian(1997)]{nulsen97}
	Nulsen, P.\ E.\ J., \& Fabian, A.\ C. 1997, \mnras, 291, 425

\bibitem[Patton et al.(2002)]{patton02}
	Patton, D.\ R., et al. 2002, \apj, 565, 208

\bibitem[Peletier et al.(1990)]{peletier90}
	Peletier, R.\ F., Davies, R.\ L., Illingworth, G.\ D., 
	Davis, L.\ E., \& Cawson, M. 1990, \aj, 100, 1091

\bibitem[Poggianti et al.(1999)]{poggianti99}
	Poggianti, B.\ M., Smail, I., Dressler, A., Couch, W.\ J., 
	Barger, A.\ J., Butcher, H., Ellis, R.\ S., \& Oemler Jr., A.
	1999, \apj, 518, 576

\bibitem[Pozzetti et al.(2003)]{pozetti03}
	Pozzetti, L., et al. 2003, \aap, 402, 837

\bibitem[Pryke et al.(2002)]{pryke02}
	Pryke, C., Halverson, N.\ W., Leitch, E.\ M., Kovac, J., 
	Carlstrom, J.\ E., Holzapfel, W.\ L., \& Dragovan, M. 2002, 
	\apj, 568, 46

\bibitem[Rix et al.(2004)]{rix04}
	Rix, H.-W., et al. 2004, \apjs, in press (astro-ph/0401427)

\bibitem[R\"oser \& Meisenheimer(1991)]{roser91}
	R\"oser, H.-J., \& Meisenheimer, K. \aap, 252, 458

\bibitem[Rusin et al.(2003)]{rusin03}
	Rusin, D., et al. 2003, \apj, 587, 143

\bibitem[Salpeter(1955)]{salpeter}
        Salpeter, E.\ E. 1955, \apj, 121, 161

\bibitem[Sandage \& Visvanathan(1978)]{sandage78}
	Sandage, A., \& Visvanathan, N. 1978, \apj, 225, 742

\bibitem[Sandage, Tammann, \& Yahil(1979)]{sty}
	Sandage, A., Tammann, G.\ A., \& Yahil, A. 1979, \apj, 232, 352

\bibitem[Schade et al.(1999)]{schade99}
	Schade, D., et al. 1999, \apj, 525, 31

\bibitem[Schechter(1976)]{schechter}
        Schechter, P. 1976, \apj, 203, 297

\bibitem[Schlegel, Finkbeiner, \& Davis(1998)]{sfd}
	Schlegel, D.\ J., Finkbeiner, D.\ P., \& Davis, M. 1998, \apj,
	500, 525

\bibitem[Schweizer \& Seitzer(1992)]{schweizer92}
	Schweizer, F., \& Seitzer, P. 1992, \aj, 104, 1039

\bibitem[Skrutskie et al.(1997)]{skrut}
	Skrutskie, M.\ F., et al. 1997, in `The Impact of 
	Large Scale Near-IR Sky Surveys', eds. F. Garzon et al., 
	p 25. (Dordrecht: Kluwer Academic Publishing Company)

\bibitem[Somerville \& Primack(1999)]{somer}
	Somerville, R.\ S., \& Primack, J.\ R. 1999, \mnras, 310, 1087

\bibitem[Somerville et al.(2004)]{somer04}
	Somerville, R.\ S., Lee, K., Ferguson, H.\ C., Gardner, J.\ P.,
	Moustakas, L.\ A., \& Giavalisco, M. 2004, \apj, 600, 171

\bibitem[Spergel et al.(2003)]{spergel03}
	Spergel, D.\ N., et al. 2003, \apjs, 148, 175

\bibitem[Stoughton et al.(2002)]{edr}
	Stoughton, C., et al. 2002, \aj, 123, 485

\bibitem[Strateva et al.(2001)]{strateva01}
	Strateva, I., et al. 2001, \aj, 122, 1861

\bibitem[Terlevich, Caldwell, \& Bower(2001)]{terlevich01}
	Terlevich, A.\ I., Caldwell, N., \& Bower, R.\ G. 2001, \mnras, 326, 
	1547

\bibitem[Toomre \& Toomre(1972)]{toomre}
        Toomre, A., \& Toomre, J.  1972, \apj, 178, 623

\bibitem[Totani \& Yoshii(1998)]{totani98}
	Totani, T., \& Yoshii, Y. 1998, \apj, 501, L177

\bibitem[Trager et al.(2000)]{trager00}
	Trager, S.\ C., Faber, S.\ M., Worthey, G., \& Gonz\'alez, J.\ J.
	2000, \aj, 120, 165

\bibitem[Treu et al.(2002)]{treu02}
	Treu, T., Stiavelli, M., Casertano, S., M{\o}ller, P., \& 
	Bertin, G. 2002, \apj, 564, L13

\bibitem[Tully et al.(1998)]{tully98}
	Tully, R.\ B., Pierce, M.\ J., Huang, J.-S., Saunders, W.,
	Verheijen, M.\ A.\ W., \& Witchalls, P.\ L. 1998, \aj, 115, 2264

\bibitem[van de Ven, van Dokkum, \& Franx(2003)]{vandeven03}
	van de Ven, G., van Dokkum, P.\ G., \& Franx, M. 2003, 
	\mnras, 344, 924

\bibitem[van Dokkum \& Franx(2001)]{vandokkum01}
	van Dokkum, P.\ G., \& Franx, M. 2001, \apj, 553, 90

\bibitem[van Dokkum et al.(2000)]{vandokkum00}
	van Dokkum, P.\ G., \& Franx, M., Fabricant, D., Illingworth, G.\
	D., \& Kelson, D.\ D. 2000, \apj, 541, 95

\bibitem[van Dokkum et al.(2001)]{vandokkum01a}
	van Dokkum, P.\ G., Stanford, S.\ A., Holden, B.\ P.,
	Eisenhardt, P.\ R., Dickinson, M., \& Elston, R. 2001, \apj, 552, L101

\bibitem[van Dokkum et al.(2003)]{vandokkum03}
	van Dokkum, P.\ G., et al. 2003, \apj, 587, L83

\bibitem[Vazdekis et al.(2001)]{vazdekis01}
	Vazdekis, A., Kuntschner, H., Davies, R.\ L., Arimoto, N.,
	Nakamura, O., \& Peletier, R. 2001, \apj, 551, L127

\bibitem[White \& Frenk(1991)]{white91}
	White, S.\ D.\ M., \& Frenk, C.\ S. 1991, \apj, 379, 52

\bibitem[Wolf et al.(2001a)]{wolf01a}
	Wolf, C., Meisenheimer, K., R\"oser, H.-J. 2001, \aap, 365, 660 (W01)

\bibitem[Wolf et al.(2001b)]{wolf01b}
	Wolf, C., et al. 2001b, \aap, 365, 681

\bibitem[Wolf et al.(2003)]{wolf03}
	Wolf, C., Meisenheimer, K., Rix, H.-W., Borch, A., Dye, S., \&
	Kleinheinrich, M. 2003, \aap, 401, 73 (W03)

\bibitem[Yan \& Thompson(2003)]{yan03}
	Yan, L., \& Thompson, D. 2003, \apj, 586, 765

\bibitem[Ziegler et al.(2001)]{ziegler01}
	Ziegler, B.\ L., Bower, R.\ G., Smail, I., Davies, R.\ L.,
	Lee, D. 2001, \mnras, 325, 1571

\end{thebibliography}
\end{document}